\documentclass[12pt]{article}
\usepackage{graphicx}
\usepackage{amsmath}
\newcommand{\sm}{\hbox{$\bigcirc$\kern-0.72em\hbox{\bf s} }}
\newcommand{\Id}{\hbox{\sl 1\kern-0.25em\hbox{I}}}

\newcommand{\rcorr}{\hbox{\kern-1.2em$\longrightarrow$}}
\newcommand{\lrcorr}{\hbox{\kern-1.2em$\longleftrightarrow$}}

\newcommand{\nRightarrow}{\Rightarrow\kern-1.2em\hbox{/}\kern.8em} %
\newcommand{\BB}{\hbox{I\kern-.2em\hbox{B}}} 
\newcommand{\DD}{\hbox{I\kern-.2em\hbox{D}}} 
\newcommand{\FF}{\hbox{I\kern-.2em\hbox{F}}} 
\newcommand{\NN}{\hbox{I\kern-.2em\hbox{N}}}  
\newcommand{\ZZ}{{{\rm Z}\kern-.28em{\rm Z}}} 
\newcommand{\RR}{\mathop{{\rm I}\kern-.2em{\rm R}}\nolimits} 
\newcommand{\RRe}{\mathop{{\rm I}\kern-.2em{\rm Re}}\nolimits} 
\newcommand{\QQ}{\hbox{l\kern-.36em\hbox{Q}}}  
\newcommand{\CC}{\hbox{{\textsf I}\kern-.47em\hbox{C}}}
\newcommand{\nop}{\hbox{{\textsf I}\kern-.47em\hbox{O}}}

\def\TREV{{{}^\triangleleft\kern-1.5pt\texttt{T}}}
\def\trev{{{}^\triangleleft\kern-3.2pt\texttt{t}}}
\def\SREV{{{}_\triangleleft\kern-2pt\texttt{S}}}
\def\srev{{{}_\triangleleft\kern-2.2pt\texttt{s}}}

\begin{document}
\title{New representations of Poincar\'e group\\
for  consistent relativistic particle theories}
\author{Giuseppe Nistic\`o
\\
{\small Dipartimento
di Matematica e Informatica, Universit\`a della Calabria, Italy}\\
{\small and}\\
{\small
INFN -- gruppo collegato di Cosenza, Italy}\\
{\small email: giuseppe.nistico@unical.it} } \maketitle
\abstract{
Though the irreducible representations of the Poincar\'e group form the groundwork for the formulation
of relativistic quantum theories of a particle,
robust classes of such representations are missed in current formulations of these theories.
In  this work the extended class of irreducible representations with positive ``mass'' parameter is explicitly determined.
Several new representations in such extension, so far excluded, give rise to consistent theories for Klein-Gordon particles and also to new species
of particle theories.
}
\section{Introduction}
The identification of the irreducible representations of the Poincar\'e group $\mathcal P$
lays the groundwork for the formulation of the relativistic quantum theories of one elementary free particle.
Indeed, each such a theory must contain
\cite{b1} an irreducible representation $g\to U_g$ of $\mathcal P$
that realizes the quantum transformation of every quantum observable according to $A\to S_g[A]=U_g A U_g^{-1}$.
Unfortunately, the literature about relativistic quantum theories of a single particle does not take into account all possible
irreducible representations of the Poincar\'e group $\mathcal P$.
One of the classes discarded is that of the irreducible representations with anti-unitary space inversion operator \cite{W2}, \cite{JM},
\cite{Costa}; in fact, not only
quantum theories of a particle characterized by anti-unitary space inversion operator can be consistently developed \cite{Gr32}, but even
anti-unitary space inversion operators turn out to be indispensable for formulating {\sl complete} quantum theories of Klein-Gordon particles without the inconsistencies
that plagued the early theory \cite{WeinBook}.
\par
These arguments point out that the following tasks should be accomplished in order to effectively identify the possible quantum theories of a free particle.
\begin{itemize}
\item[{\bf T.1.}]
To single out the possible irreducible representations of $\mathcal P$ without {\it a priori} preclusions, such as the preclusion against
representations with anti-unitary space inversion operator.
\item[{\bf T.2.}]
Then,
the explicit determination of the theories for an elementary free particle
can be addressed by selecting which of the representations identified by {\bf T.1} satisfy the further constraints
imposed by the peculiar features characterizing this specific physical system.
Only the representations inconsistent with these further constraints have to be excluded.
\end{itemize}
In this article we address the first task. In order to confer linearity to the presentation,
we carry out a general classification and identification
by means of a systematic self-contained derivation.
\par
Task {\bf T.2} is outside the scope of the present article. However, in order to ascertain that our work is not
meaningless from a {\sl theoretical physics} point of view,
in the final section 6 we show that consistent relativistic quantum theories of a particle can be formulated, which are based on
irreducible representations singled out by the present work and not considered in the literature.
\vskip.5pc
Section 2 introduces the notation and basic mathematical prerequisites relative to Poincar\'e groups and their representations.
\vskip.4pc
In section 3 we show that all irreducible representations of $\mathcal P$ can be classified according to the following three criteria.
\vskip.3pc\noindent
{\sl ``Mass'' and ``spin'' parameters $(\mu,s)$}.
\quad Each irreducible representation of $\mathcal P$ must be characterized by a unique pair $(\mu,s)$ , $\mu\in\CC$, $s\in\frac{1}{2}\NN$,
called {\sl mass} and {\sl spin} parameters, respectively. In this work we restrict to the class of
positive {\sl mass} irreducible representations.
\vskip.3pc\noindent
{\sl Spectrum of $P_0$}.
\quad In every irreducible representation there are only three mutually exclusive possibilities
for the spectrum $\sigma(P_0)$ of the Hamiltonian operator $P_0$:
\vskip.2pc
{\sl Either} $\sigma(P_0)=[\mu,\infty)\equiv{\mathcal I}_\mu^+$; \vskip.2pc
{\sl or} $\sigma(P_0)=(-\infty,-\mu]\equiv{\mathcal I}_\mu^-$;
\vskip.2pc
{\sl or} $\sigma(P_0)={\mathcal I}_\mu^-\cup{\mathcal I}_\mu^+$.
\vskip.2pc\noindent
It is shown how these possibilities are related to the unitary or anti-unitary character of the space inversion operator $\SREV$ and of the
time reversal operator $\TREV$.
\vskip.3pc\noindent
{\sl Reducibility of $U^\pm$}.
\quad Given an irreducible representation $U$ of $\mathcal P$ characterized by a pair $(\mu,s)$ and by one of the possible spectra of $P_0$,
it turns out that two particular sub-representations, $U^+$ or $U^-$,
can be reducible or not. The literature takes into account only irreducible representations with $U^\pm$ irreducible.
Our redetermination does not overlook the irreducible representations with $U^+$ or
$U^-$ reducible.
\vskip.4pc
In section 4 we explicitly identify all irreducible representations of $\mathcal P$ with $U^\pm$ irreducible.
The representations with $\sigma(P_0)=I_\mu^+$ or $\sigma(P_0)=I_\mu^-$ are already well known. For
$\sigma(P_0)=I_\mu^+\cup I_\mu^-$ we identify, besides the well known representations with both
space inversion operator $\SREV$ and time reversal operator $\TREV$ unitary, also all irreducible representations
with $\SREV$ anti-unitary and $\TREV$ unitary, and with both $\SREV$, $\TREV$ anti-unitary, neglected in the literature.
\vskip.4pc
Section 5 deals with the class of the so far ``ignored'' representations of $\mathcal P$ with $U^+$
or $U^-$ reducible. We explicitly
identify irreducible representations $U$ of $\mathcal P$ with $U^\pm$ reducible in all three possible cases
with $\sigma(P_0)=I_\mu^+$, $\sigma(P_0)=I_\mu^-$ and $\sigma(P_0)=I_\mu^+\cup I_\mu^-$.
They open to the possibility of yet unknown  particle theories. Section 6 shows that this possibility is absolutely concrete.
\section{Notation and mathematical prerequisites}
\subsection{Poincar\'e group.}
Given any vector $\underline x=(x_0,x_1,x_2,x_3)\equiv(x_0,{\bf x})\in\RR^4$, we call $x_0$ the {\sl time component} of $\underline x$
and ${\bf x}=(x_1,x_2,x_3)$ the {\sl spatial component} of $\underline x$.
The {\sl proper orthochronous} Poincar\'e group ${\mathcal P}_+^\uparrow$ is the separable locally compact
group of all transformations of $\RR^4$
generated by the ten one-parameter sub-groups ${\mathcal T}_0$, ${\mathcal T}_j,{\mathcal R}_j$, ${\mathcal B}_j$, $j=1,2,3$,
of time translations, spatial translation, proper spatial rotations and Lorentz boosts, respectively.
The Euclidean group $\mathcal E$ is the sub-group generated by all ${\mathcal T}_j$ and ${\mathcal R}_j$.
The sub-group generated by
${\mathcal R}_j$, ${\mathcal B}_j$ is the proper orthochronous Lorentz group ${\mathcal L}_+^\uparrow$ \cite{BarBook};
it does not include time reversal $\trev$ and space inversion $\srev$.
Time reversal $\trev$ transforms $\underline x=(x_0,{\bf x})$ into $(-x_0,{\bf x})$; space inversion $\srev$
transforms $\underline x=(x_0,{\bf x})$ into $(x_0,-{\bf x})$.
The group generated by $\{{\mathcal P}_+^\uparrow, \trev,\srev\}$ is the separable and locally compact Poincar\'e group $\mathcal P$.
By ${\mathcal L}_+$ we denote the subgroup generated by ${\mathcal L}_+^\uparrow$ and $\trev$, while ${\mathcal L}^\uparrow$
denotes the subgroup generated by ${\mathcal L}_+^\uparrow$ and $\srev$; analogously, ${\mathcal P}_+$ denotes
the subgroup generated by ${\mathcal P}_+^\uparrow$ and $\trev$, while ${\mathcal P}^\uparrow$ is the subgroup
generated by ${\mathcal P}_+^\uparrow$ and $\srev$.
\subsection{Mathematical structures.}
The following mathematical structures, based on a complex and separable Hilbert space $\mathcal H$,
are of general interest in quantum theory.
\begin{description}
\item[\quad-]
The set $\Omega({\mathcal H})$ of all self-adjoint operators of $\mathcal H$; in a quantum theory these operators represent
{\sl quantum observables}.
\item[\quad-]
The lattice $\Pi({\mathcal H})$ of all projections operators of $\mathcal H$; in a quantum theory they represent observables with spectrum $\{0,1\}$.
\item[\quad-]
The set $\Pi_1({\mathcal H})$ of all rank one orthogonal projections of $\mathcal H$.
\item[\quad-]
The set ${\mathcal S}({\mathcal H})$ of all density operators of $\mathcal H$;
in a quantum theory these operators represent {\sl quantum states}.
\item[\quad-]
The set ${\mathcal V}({\mathcal H})$ of all unitary  or anti-unitary operators of the Hilbert space $\mathcal H$.
\item[\quad-]
The set ${\mathcal U}({\mathcal H})$ of all unitary operators of $\mathcal H$; trivially,
${\mathcal U}({\mathcal H})\subseteq{\mathcal V}({\mathcal H})$ holds.
\end{description}
\subsection{Generalized representations of groups.}
The following definition introduces generalized notions of group representation.
\vskip.5pc\noindent
{\bf Definition 2.1.} {\sl Let $G$ be a separable, locally compact group with identity element $e$. A correspondence
$U:G\to{\mathcal V}({\mathcal H})$, $g\to U_g$, with $U_e=\Id$, is a generalized projective representation of $G$
if the following conditions are satisfied.
\begin{description}
\item[\;{\rm i)}\;\;]
A complex function $\sigma:{G}\times{G}\to{\CC}$,
called multiplier,
exists such that $U_{g_1g_2}=\sigma(g_1,g_2)U_{g_1}U_{g_2}$; the modulus $\vert\sigma(g_1,g_2)\vert$ is always 1, of course;
\item[\;{\rm ii)}\;]
for all $\phi,\psi\in\mathcal H$, the mapping $g\to\langle U_g\phi\mid\psi\rangle$ is a Borel function in $g$.
\end{description}\noindent
If $U_g$ is unitary for all $g\in G$, then $U$ is called
projective representation.
\par\noindent
A generalized projective representation is said to be continuous if for any fixed $\psi\in\mathcal H$
the mapping $g\to U_g\psi$ from $G$ to $\mathcal H$ is continuous with respect to $g$.
}
\vskip.5pc\noindent
If $g\to U_g$ is a generalized projective representation of $\mathcal P$ and $\theta(g)\in\RR$, then $g\to \tilde U_g=e^{i\theta(g)}U_g$
is a generalized projective representation, said {\sl equivalent} \cite{b3} to $g\to U_g$.
\vskip.4pc\noindent
In \cite{N1} we have proved that the following statement holds.
\vskip.5pc\noindent
{\bf Proposition 2.1.}
{\sl If $G$ is a connected group, then every continuous generalized projective representation of $G$ is a projective
representation, i.e. $U_g\in{\mathcal U}({\mathcal H})$, for all $g\in G$.}
\subsection{Generalized representations of the Poincar\'e group $\mathcal P$}
All sub-groups ${\mathcal T}_0$, ${\mathcal T}_j,{\mathcal R}_j$, ${\mathcal B}_j$ of ${\mathcal P}_+^\uparrow$ are additive;
in fact, ${\mathcal B}_j$ is not additive with respect to the parameter
{\sl relative velocity} $u$, but it is additive with respect to the parameter $\varphi(u)=\frac{1}{2}\ln\frac{1+u}{1-u}$.
Then, according to Stone's theorem \cite{c9}, for every continuous projective representation of ${\mathcal P}_+^\uparrow$, an equivalent projective representation $U$ exists for which there are
ten self-adjoint generators $P_0$, $P_j$, $J_j$, $K_j$, $j=1,2,3$, of the ten
one-parameter unitary subgroups $\{e^{iP_0t}\}$,   $\{e^{-i{P}_j a_j},\,a\in{\RR}\}$,
$\{e^{-i{J}_j \theta_j},\,\theta_j\in{\RR}\}$, $\{e^{i{K}_j \varphi(u_j)},\,u_j\in{\RR}\}$ of ${\mathcal U}({\mathcal H})$
that represent the one-parameter
sub-groups ${\mathcal T}_0$, ${\mathcal T}_j,{\mathcal R}_j$, ${\mathcal B}_j$ according to $U$.
\subsubsection{Commutation relations.}
The
structural properties of ${\mathcal P}_+^\uparrow$ as a Lie group imply that
every continuous projective representation of ${\mathcal P}_+^\uparrow$ admits an equivalent projective representation $U$ such that
the following commutation relations \cite{c13} hold for its generators.
\vskip.5pc\noindent
(i)\hskip4.3mm $[{P}_j,{P}_k]=\nop$,\hskip15mm (ii) $[{J}_j,{P}_k]=i{\hat\epsilon}_{jkl}{P}_l$,
\hskip8mm (iii) $[{ J}_j,{ J}_k]=i{\hat\epsilon}_{jkl}{ J}_l$,\par\noindent
(iv)\hskip2.3mm $[{ J}_j,{ K}_k]
=i{\hat\epsilon}_{jkl}{ K}_l$,\quad\hskip1.8mm (v)\hskip1mm $[{ K}_j,{ K}_k]=-i\delta_{j,k}J_l$,
\quad (vi) $[{ K}_j,{ P}_k]=i\delta_{jk}P_0$,\hfill{}(1)
\par\noindent
(vii) $[P_j,P_0]=\nop$,\hskip15.2mm (viii) $[J_j,P_0]=\nop$,\hskip14.8mm (ix) $[K_j,P_0]=iP_0$,
\vskip.4pc\noindent
where ${\hat\epsilon}_{jkl}$ is the Levi-Civita symbol ${\epsilon}_{jkl}$
restricted by the condition $j\neq l\neq k$.
\par
Let $U:{\mathcal P}\to{\mathcal V}({\mathcal H})$ be a generalized projective representation of $\mathcal P$, whose restriction to ${\mathcal P}_+^\uparrow$ is continuous.
Since time reversal
$\trev$ and space inversion $\srev$ are {\sl not connected} with the identity transformation $e\in\mathcal P$,
the operators $\TREV=U_\trev$ and $\SREV=U_\srev$ can be unitary or anti-unitary.
The phase factor $e^{i\theta(g)}$ can be always chosen \cite{c13} in such a way that the following statements hold in the equivalent generalized projective representation.
\vskip.5pc\noindent
{\sl If $\SREV$ {\sl is unitary},
\par\noindent
then $[\SREV,P_0]=\nop$,\quad $\SREV P_j=-P_j\SREV$,\quad $[\SREV,J_k]=\nop$,\quad $\SREV K_j=-K_j\SREV$;}\quad
$\SREV^2=\Id$;\hfill{(2)}
\vskip.4pc\noindent
{\sl If $\SREV$ is anti-unitary,
\par\noindent
then $\SREV P_0=-P_0\SREV$,\quad $[\SREV, P_j]=\nop$,\quad $\SREV J_k=-J_k\SREV$,\quad $\SREV K_j=K_j\SREV$,}\hfill{(3)}\par
{\sl and $\SREV^2=c\Id$, so that $\SREV^{-1}=c\SREV$, where $c=1$ or $c=-1$.}
\vskip.5pc\noindent
{\sl If $\TREV$ is unitary,
\par\noindent
then $\TREV P_0=-P_0\TREV$,\quad $[\TREV,P_j]=\nop$,\quad $[\TREV,J_k]=\nop$,\quad $\TREV K_j=-K_j\TREV$;}\quad$\TREV^2=\Id$\hfill{(4)}
\vskip.5pc\noindent
{\sl If $\TREV$ is anti-unitary,
\par\noindent
then $\TREV P_0=P_0\TREV$,\quad $\TREV P_j=-P_j\TREV$,\quad $\TREV J_k=-J_k\TREV$,\quad $\TREV K_j=K_j\TREV$,}\hfill{(5)}\par
{\sl and $\TREV^2=c\Id$, so that $\TREV^{-1}=c\TREV$, either $c=1$ or $c=-1$ must hold.}
\vskip.5pc\noindent
{\sl The commutation condition for the pair $\SREV$,$\TREV$, is
\par\quad $\SREV\TREV=\omega\TREV\SREV$, \quad with $\omega\in\CC$ and $\vert\omega\vert=1$}.\hfill{(6)}
\vskip.5pc\noindent
From now on the continuity hypothesis for $U\mid_{{\mathcal P}_+^\uparrow}$ is implicitly assumed.
A quantum theory based on a generalized projective representation is indistinguishable in all respects from the theory based on an equivalent
$e^{i\theta}U$. For this reason we assume that a generalized projective representation of $\mathcal P$ satisfies (1)-(6).
\vskip.5pc\noindent
{\bf Proposition 2.2.}
{\sl
If $U:{\mathcal P}\to{\mathcal V}({\mathcal H})$ is a generalized projective representation, then
the relations (1)-(6) imply that the following equalities hold for all $g\in{{\mathcal P}}$, including $\trev$ and $\srev$.
$$
[P_0^2-(P_1^2+P_2^2+P_3^2),U_g]=\nop\,,\eqno(7)
$$
$$
[W_0^2
-(W_1^2+W_2^2+W_3^2),U_g]=\nop\,,\eqno(8)
$$
where $W_0={\bf P}\cdot{\bf J}$ and $W_j=P_0J_j-({\bf P}\times{\bf K})_j$ define the Luba\'nski four-operator $\underline W=(W_0, {\bf W})$.
}
\subsection{Spectral properties of the self-adjoint generators}
Spectral properties of the self-adjoint generators are now derived.
Relations (1.i), (1.vii) establish that the generators $P_0$, $P_1$, $P_2$, $P_3$ of a generalized projective representation $U$ of $\mathcal P$
commute with each other; therefore,
according to spectral theory \cite{RS} a common spectral measure
$E:{\mathcal B}({\RR}^4)\to\Pi({\mathcal H})$
exists such that
$$
P_0=\int\lambda dE^{(0)}_\lambda\,,\quad P_j=\int\lambda dE^{(j)}_\lambda\,,\; j=1,2,3,\eqno(9)
$$
where\quad $E^{(0)}_\lambda=E((-\infty,\lambda]\times{\RR}^3)$,\quad
$E^{(1)}_\lambda=E(\RR\times(-\infty,\lambda]\times{\RR}^2)$,\quad
$E^{(2)}_\lambda=E({\RR}^2\times(-\infty,\lambda]\times{\RR})$,\quad
$E^{(3)}_\lambda=E({\RR}^3\times(-\infty,\lambda])$\quad
are the resolutions of the identity of the individual operators $P_0$, $P_1$, $P_2$, $P_3$.
\par
Once introduced the {\sl four-operator}
$\underbar P=(P_0,P_1,P_2,P_3)\equiv (P_0,{\bf P})$, the equalities (9) can be rewritten in the more compact form
$$
\underline P =\int \underline p\,dE{\underline p}\,,\hbox{ with }
dE_{\underline p}=dE^{(0)}_{p_0}dE^{(1)}_{p_1}dE^{(2)}_{p_2}dE^{(3)}_{p_3}\,.\eqno(10)
$$
The {\sl spectrum} of $\underbar P$ can be defined as the following closed subset of $\RR^4$.
$$
\sigma({\underbar P})=\{\underbar p=(p_0,{\bf p})\in\RR^4\mid
E(\Delta_{\underbar p})\neq\nop\hbox{ for every neighborough }\Delta_{\underbar p}\hbox{ of }
\underbar p\}\,.\eqno(11)
$$
By making use of (1), the following proposition can be proved.
\vskip.4pc\noindent
{\bf Proposition 2.3.}
{\sl
Let $U:{\mathcal P}\to{\mathcal U}({\mathcal H})$ be
a projective representation of ${\mathcal P}_+^\uparrow$,
Then for every Lorentz transformation $g\in{\mathcal L}_+^\uparrow$ the following relation holds
$$
U_g\underbar PU_g^{-1}=\hbox{\rm\textsf g}(\underbar P),\eqno(12)
$$
where ${\rm\textsf g}:\RR^4\to\RR^4$ is the function that transforms any
$\underbar p\in\RR^4$ as a four-vector according to $g$. Moreover, the following statement is a straightforward implication of (12).
$$
U_gE(\Delta)U_g^{-1}=E({\textsf g}^{-1}(\Delta))\hbox{ holds for every }g\in{\mathcal L}_+^\uparrow.
\eqno(13)
$$
}
\section{Classification of positive ``mass'' irreducible representations of $\mathcal P$}
A generalized projective representation $U:{\mathcal P}\to{\mathcal V}({\mathcal H})$
can be reducible or not; in the case that
it is reducible, however, it must be the direct sum or the direct integral of irreducible ones \cite{BarBook}.
Therefore, to determine all possible generalized projective representations of ${\mathcal P}$
it is sufficient to identify the irreducible ones. For this reason, from now on we specialize to
{\sl irreducible} generalized projective representations of $\mathcal P$.
Hence, from Prop. 2.2 the following proposition follows.
\vskip.4pc\noindent
{\bf Proposition 3.1.}
{\sl
If a generalized projective representation of $\mathcal P$ is irreducible,
then two real numbers $\eta$, $\varpi$ exist such that the following equalities hold.
$$
(i)\;\;P_0^2-{\bf P}^2=\eta\Id\quad\hbox{and}\quad(ii)\;\;W^2\equiv W_0^2
-(W_1^2+W_2^2+W_3^2)=\varpi\Id\,.\eqno(14)
$$}
Therefore every irreducible generalized projective representation of ${\mathcal P}$ is characterized by the real constants
$\eta,\varpi$. We restrict our investigation to those irreducible generalized representations
for which $\eta>0$, so that $\eta=\mu^2$, with $\mu>0$; with this restriction it can be proved that $s\in\frac{1}{2}\NN$ exists such that $\varpi=-\mu^2 s(s+1)$.
The parameters $\mu$ and $s$ are called {\sl mass} and {\sl spin} parameters, respectively.
\subsection{Spectral characterization of positive ``mass'' irreducible representations of $\mathcal P$ }
Now we show that for an irreducible generalized projective representation of ${\mathcal P}$, characterized by specific
parameters $\mu>0$ and $s$,
the spectrum $\sigma(\underbar P)$ of the four-operator $\underbar P=(P_0,{\bf P})$,
must be one of three definite subsets $S^+_\mu$, $S^-_\mu$, $S^+_\mu\cup S^-_\mu$ of $\RR^4$, where
$$
S^{+}_{\mu}=\{\underbar p\mid p_0^2-{\bf p}^2=\mu^2,\,p_0>0\}\,,\hbox{ (positive hyperboloid)}\,,\eqno(15)
$$
$$
S^{-}_{\mu}=\{\underbar p\mid p_0^2-{\bf p}^2=\mu^2,\,p_0<0\}\,,\hbox{ (negative hyperboloid)}\,.
$$
\vskip.4pc\noindent
{\bf Proposition 3.2.}
{\sl
If $U:{\mathcal P}\to {\mathcal V}({\mathcal H})$ is an irreducible generalized projective representation,
then there are only the following mutually exclusive possibilities for the spectra $\sigma(\underbar P)$ and $\sigma(P_0)$.
\vskip.3pc\noindent
({\bf u})\quad $\sigma(\underbar P)=S^{+}_{\mu}$ and $\sigma(P_0)=[\mu,\infty)\equiv{\mathcal I}_\mu^+$, ``up'' spectrum;
\vskip.3pc\noindent
({\bf d})\quad $\sigma(\underbar P)=S^{-}_{\mu}$ and $\sigma(P_0)=(-\infty,-\mu]\equiv{\mathcal I}_\mu^-$, ``down'' spectrum;
\vskip.3pc\noindent
({\bf s})\quad $\sigma(\underbar P)=S^{+}_{\mu}\cup S^{-}_{\mu}$ and $\sigma(P_0)={\mathcal I}_\mu^+\cup{\mathcal I}_\mu^-$, ``symmetrical'' spectrum.
}
\vskip.4pc\noindent
{\bf Proof.}
Since $P_0^2-{\bf P}^2-\mu^2=\nop$, if $\underbar p\in\sigma(\underbar P)$ then $p_0^2-{\bf p}^2-\mu^2=0$ must hold,
i.e.
$$\sigma(\underbar P)\subseteq\{\underline p\mid p_0^2-{\bf p}^2-\mu^2=0\}\equiv S^{+}_{\mu}\cup S^{-}_{\mu}\,.\eqno(16)$$
On the other hand,
if $\underbar p\in\sigma(\underbar P)$, then according to spectral theory $\textsf g(\underbar p)\in\sigma(\textsf g(\underbar P))$ holds
for all $g\in{\mathcal L}_+^\uparrow$,
of course; but $\textsf g(\underbar P)=U_g\underbar PU_g^{-1}$ by Prop. 2.3; therefore,
$\underbar p\in\sigma(\underbar P)$ if and only if $\textsf g(\underbar p)\in\sigma(\underbar P)$
because $\underbar P$ and $U_g\underbar PU_g^{-1}$ have the same spectra.
Hence, the following statements hold.
$$
\sigma(\underbar P)\cap S^{+}_{\mu}\neq\emptyset\hbox{ implies }
S^{+}_{\mu}\subseteq\sigma(\underbar P)\,;\;\;
\sigma(\underbar P)\cap S^{-}_{\mu}\neq\emptyset\hbox{ implies }
S^{-}_{\mu,\varpi}\subseteq\sigma(\underbar P)\,
.\eqno(17)$$
Since $\sigma(\underbar P)\neq\emptyset$, (16) and (17) imply that only one of the three cases
({\bf u}), ({\bf d}) or ({\bf s}) can occur.
\hfill{$\bullet$}
\vskip.5pc\noindent
In the case ({\bf s}) the restriction
$U:{\mathcal P}_+^\uparrow\to{\mathcal U}({\mathcal H})$ is always reducible, namely $U\mid_{{\mathcal P}_+^\uparrow}$ is reduced by
the projection operators $E^{+}=\int_{S^+_\mu}dE_{\underline p}\equiv\int_\mu^\infty p_0dE_{p_0}^{(0)}$
and $E^{-}=\int_{S^-_\mu}dE_{\underline p}\equiv\int_{-\infty}^{-\mu} p_0dE_{p_0}^{(0)}$, with
ranges ${\mathcal M}^+=E^+{\mathcal H}$ and ${\mathcal M}^-=E^-{\mathcal H}$, respectively.
We prove this statement in the following Proposition.
\vskip.5pc\noindent
{\bf Proposition 3.3.}
{\sl
In an irreducible generalized projective representation $U:{\mathcal P}\to{\mathcal V}({\mathcal H})$
the relation $[E^\pm,U_g]=\nop$ holds for all $g\in{\mathcal P}_+^\uparrow$.
Hence, the following consequences can be immediately implied.
\begin{itemize}
\item[i)]
In the case of symmetrical spectrum $\sigma(\underline P)=S_\mu^+\cup S_\mu^-$,
the restriction $U\mid_{{\mathcal P}_+^\uparrow}$ is reduced by $E^+$ into
$U^+=E^+U\mid_{{\mathcal P}_+^\uparrow}E^+$
\;and \;$U^-=E^-U\mid_{{\mathcal P}_+^\uparrow}E^-$;
\item[ii)]
if $\sigma(\underline P)=S_\mu^+$ (resp., $\sigma(\underline P)=S_\mu^-$),
then $U\mid_{{\mathcal P}_+^\uparrow}=U^+$ (resp., $U\mid_{{\mathcal P}_+^\uparrow}=U^-$).
\end{itemize}
}
\vskip.4pc\noindent
{\bf Proof.}
In the case ({\bf u}) and ({\bf d}), the statement is trivial because $E^\pm=\int_{S_\mu^\pm}dE_{\underline p}\equiv\int_{\sigma(\underline P)}dE_{\underline p}=\Id$.
Then we suppose that $\sigma(\underline P)=S_\mu^+\cup S_\mu^-$.
Since $E^+=\chi_{S^+_\mu}(\underline P)$,
where $\chi_{S^+_\mu}$ is the characteristic functional of $S^+_\mu$, the relations (1.i),(1.vii) imply that
$E^+$ commutes with $P_0$ and with all $P_j$. Therefore it remains to show that ${\mathcal M}^+$ is left invariant by
$U_g$, for every $g\in{\mathcal L}_+^\uparrow$.
If $\psi\in{\mathcal M}^+$, then for every $g\in{\mathcal L}_+^\uparrow$ we have
$U_g\psi=U_g\int_{S^+_\mu}dE_{\underline p}\psi=
\int_{{S^+_\mu}}dU_gE_{\underline p}{U_g}^{-1}(U_g\psi)
=\int_{S^+_\mu}dE_{{\textsf g}^{-1}({\underline p})}(U_g\psi)$, by Prop. 2.3.
The last integral is a vector of ${\mathcal M}^+=
E^+\mathcal H$, because ${\textsf g}^{-1}(\underline p)\in S_\mu^+$ if $\underline p\in S_\mu^+$ for $g\in{\mathcal L}_+^\uparrow$.
\par\noindent
The same argument, suitably adapted, proves that ${\mathcal M}^-$ is left invariant by $U_g$, for every $g\in{\mathcal P}_+^\uparrow$.
The consequences (i) and (ii) are straightforward.\hfill{$\bullet$}
\subsubsection{Spectral implications of the unitary or anti-unitary character of $\SREV$ and $\TREV$}
Now we show how each of the possibilities for $\sigma(\underline P)$ established by Prop.3.2 is characterizable
according to the unitary or anti-unitary character of the time reversal and the space inversion operators
$\TREV$ and $\SREV$.
\vskip.5pc\noindent
{\bf Lemma 3.1.}
{\sl Let $T$ be a unitary or anti-unitary operator, and let $A$ be a self-adjoint operator with spectral measure
$E^A:{\mathcal B}(\RR)\to \Pi({\mathcal H})$. If
$TAT^{-1}=f(A)$, where $f$ is a continuous bijection of $\RR$, then
$TE^A(\Delta)T^{-1}=E^A(f^{-1}(\Delta))$, for all $\Delta\in{\mathcal B}(\RR)$.}
\vskip.4pc\noindent
{\bf Proof.}
We recall that if $T$ is unitary or anti-unitary, then an operator $D$ is a projection operator if and only if $TDT^{-1}$ is a projection operator;
moreover, if $\Delta\to E^A(\Delta)$ is the spectral measure of $A$, then $\Delta\to F(\Delta)=TE_A(\Delta)T^{-1}$ is
the spectral measure of $f(A)$. Now, let us define $\tilde\Delta=f^{-1}(\Delta)$. If $\pi(-a,a)$ is a partition of the interval
$[-a,a]$ formed by sub-intervals $\tilde\Delta_j$ with $\tilde\lambda_j\in\tilde\Delta_j$, then according to spectral theory we can write
\par\noindent
$f(A)=\lim_{{\Vert\pi(-a,a)\Vert\to 0}\atop {a\to\infty}}\sum_j f(\tilde\lambda_j)E(\tilde\Delta_j)=
\lim_{{\Vert\pi(-a,a)\Vert\to 0}\atop {a\to\infty}}\sum_j \lambda_j E(f^{-1}(\Delta_j))$\par\quad
$=\lim_{{\Vert\pi(-a,a)\Vert\to 0}\atop {a\to\infty}}\sum_j \lambda_j F(\Delta_j)$, where $\lambda_j=f(\tilde\lambda_j)\in\Delta_j$.
\par\noindent
Therefore, for the uniqueness of the spectral measure $F$ of $f(A)$ we have
\par\noindent
$F(\Delta)=E(f^{-1}(\Delta))=TE(\Delta)T^{-1}$.
\hfill{$\bullet$}
\vskip.5pc\noindent
{\bf Proposition 3.4.}
Let $U:{\mathcal P}\to{\mathcal U}({\mathcal H})$ be an irreducible generalized projective representation.
{\sl If {$\TREV$} is anti-unitary and {\rm $\SREV$} is unitary, then either
$\sigma(\underbar P)=S_\mu^+$ or $\sigma(\underbar P)=S_\mu^-$, and hence $\sigma(\underbar P)=S_\mu^+\cup S_\mu^-$
cannot occur.}
\vskip.4pc\noindent
{\bf Proof.}
First we show that the hypotheses imply that ${\mathcal M}^+$ and ${\mathcal M}^-$ are invariant under both $\TREV$ and $\SREV$.
According to (5) the relation $\TREV P_0\TREV^{-1}=P_0$ holds when $\TREV$ is anti-unitary;
therefore Lemma 3.1 applies with $A=P_0$, $T=\TREV$ and $f$ the identity function, so that
$\TREV E^{(0)}(\Delta)\TREV^{-1}=E^{(0)}(\Delta)$ holds. This implies that
if $\psi$ is any non vanishing vector in ${\mathcal M}^+$, then
$\TREV\psi=\int_\mu^\infty d(\TREV E_{p_0}^{(0)}\TREV^{-1})\TREV\psi
=\int_\mu^\infty d\,E^{(0)}_{p_0}\TREV\psi\equiv E^+\TREV\psi$. Thus $\TREV\psi$ is a vector in ${\mathcal M}^+$.
This argument can be repeated with $\SREV$ instead of $\TREV$, to deduce, by making use of (2), that $\SREV\psi\in{\mathcal M}^+$
for all $\psi\in{\mathcal M}^+$.
The invariance of ${\mathcal M}^{-}$ is proved quite similarly.
\par
Now, since ${\mathcal M}^+$ and ${\mathcal M}^-$ are invariant under the restriction $U\mid_{{\mathcal P}_+^\uparrow}$ according to Prop. 3.2,
they are invariant under the whole $U$.
If $\sigma(\underline P)=S_\mu^+\cup S_\mu^-$ held, then ${\mathcal M}^+$ would be a {\sl proper}
subspace of $\mathcal H$, so that $U$ would be reducible, in contradiction with the hypothesis of irreducibility.
\hfill{$\bullet$}
\vskip.5pc\noindent
{\bf Proposition 3.5.}
{\sl If $\TREV$ is unitary then $\sigma(\underline P)=
S_\mu^+\cup S_\mu^-$ holds, independently of $\SREV$.}
\vskip.4pc\noindent
{\bf Proof.}
If $\TREV$ is unitary, then
$\sigma(\TREV P_0\TREV^{-1})=\sigma(P_0)$. But $\TREV P_0\TREV^{-1}=-P_0$ holds by (4);
therefore $-\sigma(P_0)\equiv\sigma(-P_0)=\sigma(\TREV P_0\TREV^{-1})=\sigma(P_0)$, i.e.
$$
\sigma(P_0)=-\sigma(P_0).\eqno(18)
$$
Now,
in general we have $\sigma(\underline P)=S_\mu^+$ if and only if $\sigma(P_0)=[\mu,\infty)$,
$\sigma(\underline P)=S_\mu^-$ if and only if $\sigma(P_0)=(-\infty,-\mu]$, and
$\sigma(\underline P)=S_\mu^+\cup S_\mu^-$ if and only if $\sigma(P_0)=(-\infty,-\mu]\cup[\mu,\infty)$;
equation (18) holds only if $\sigma(P_0)=(-\infty,-\mu]\cup[\mu,\infty)$; thus
$\sigma(\underline P)=S_\mu^+\cup S_\mu^-$.\hfill{$\bullet$}
\vskip.5pc\noindent
{\bf Proposition 3.6.}
{\sl
If $\SREV$ is anti unitary then $\sigma(\underline P)=S_\mu^+\cup S_\mu^-$, independently of $\TREV$.}
\vskip.4pc\noindent
{\bf Proof.}
Since $\sigma(P_0)$ is not empty, at least one of the projection operators
$E^+=E^{(0)}([\mu,\infty))$ or $E^-=E^{(0)}((-\infty,-\mu])$
must be different from the null operator\; $\nop$.
Let us suppose that $E^{(0)}([\mu,\infty))\neq\nop$, so that $S_\mu^+\subseteq\sigma(\underline P)$. Since
$\SREV P_0\SREV^{-1}=-P_0$ holds by (3), Lemma 3.1 can be applied to deduce
$\SREV E^{(0)}([\mu,\infty))\SREV^{-1}=E^{(0)}((-\infty,-\mu])$; hence $E^{(0)}((-\infty,-\mu])$ is a non null projection operator
because $E^{(0)}([\mu,\infty))$ is non-null and $\SREV$ is anti-unitary. This means that
$\sigma(P_0)\cap(-\infty,-\mu]$ is not empty, that is to say
that $\sigma(\underline P)\cap S_\mu^-\neq\emptyset$; therefore, according to Prop. 3.2,
$\sigma(\underline P)=S_\mu\cup S_\mu^-$.
\par
In the case $E^{(0)}((-\infty,-\mu])\neq\nop$ the argument is easily adapted to
reach the same conclusion.\hfill{$\bullet$}
\vskip.5pc\noindent
The following proposition is an easy corollary of these results
\vskip.5pc\noindent
{\bf Proposition 3.7.}
{\sl
If $\sigma(\underline P)=S^{+}_{\mu}\cup S^{-}_{\mu}$ every vector $\psi\in\mathcal H$ can be represented
as a column vector $\psi\equiv\left[\begin{array}{c}\psi^+\cr \psi^-\end{array}\right]$, where $\psi^+=E^+\psi$ and
$\psi^-=E^-\psi$. Coherently with such a representation, any linear or anti-linear operator $A$ is represented by a matrix
$A\equiv\left[\begin{array}{cc}A_{11}&A_{12}\cr A_{21}&A_{22}\end{array}\right]$, where $A_{11}=E^+AE^+$, $A_{1}=E^+AE^-$, $A_{21}=E^-AE^+$
and $A_{22}=E^-AE^-$, in such a way that $A\psi=\left[\begin{array}{cc}A_{11}&A_{12}\cr A_{21}&A_{22}\end{array}\right]\left[\begin{array}{c}\psi^+\cr \psi^-\end{array}\right]$.
Since $[E^\pm, P_0]=[E^\pm,P_j]=[E^\pm,J_j]=[E^\pm,K_j]=\nop$,
the generators $P_0$, $P_j$, $J_k$, $K_j$ have a diagonal form.
\par
The operators $\SREV$ and $\TREV$ have diagonal representation only if are unitary and anti-unitary, respectively.}
\subsection{General classification}
An effective help, in explicitly identifying the possible structures of the irreducible generalized projective representations
of the Poincar\'e group, will be provided just by the
investigation of the reductions $U^+$ or $U^-$ singled out by Prop. 3.3.
In general, even if the ``mother'' irreducible generalized projective representation $U$ is irreducible, the reductions
$U^+$ or $U^-$
can be reducible or not.
Let us denote the class of all irreducible generalized projective representations of $\mathcal P$ by ${\mathcal I}_{\mathcal P}$
(unitarily equivalent representations are identified in ${\mathcal I}_{\mathcal P}$).
In virtue of Prop. 3.1, we can operate a classification of the representations in ${\mathcal I}_{\mathcal P}$ according to the characterizing parameters $\mu$ and $s$:
\vskip.5pc\noindent
{\bf C.1.}\quad ${\mathcal I}_{\mathcal P}=\cup_{\mu>0, s\in\frac{1}{2}\NN}\;{\mathcal I}_{\mathcal P}(\mu,s)$,
\vskip.5pc\noindent
where ${\mathcal I}_{\mathcal P}(\mu,s)$ is the class of all representations in
${\mathcal I}_{\mathcal P}$ such that $P_0^2-{\bf P}^2=\mu^2\Id$ and $W^2=-\mu^2 s(s+1)\Id$.
In its turn, by Prop. 3.2, each class ${\mathcal I}_{\mathcal P}(\mu,s)$ in {\bf C.1},  can be decomposed as
\vskip.5pc\noindent
{\bf C.2.}\quad ${\mathcal I}_{\mathcal P}(\mu,s)=
{\mathcal I}_{\mathcal P}(S_\mu^+,s)\cup {\mathcal I}_{\mathcal P}(S_\mu^-,s)\cup{\mathcal I}_{\mathcal P}(S_\mu^+\cup S_\mu^-,s)$,
\vskip.5pc\noindent
where ${\mathcal I}_{\mathcal P}(S_\mu^\pm,s)$ is the class of all representations in ${\mathcal I}_{\mathcal P}(\mu,s)$ such that
$\sigma(\underline P)=S_\mu^\pm$, and
${\mathcal I}_{\mathcal P}(S_\mu^+\cup S_\mu^-,s)$ is the class of all representations in
${\mathcal I}_{\mathcal P}(\mu,s)$ such that $\sigma(\underline P)=S_\mu^+\cup S_\mu^-$.
Each component of ${\mathcal I}_{\mathcal P}(\mu,s)$ in {\bf C.2} can be further decomposed
into two sub-classes according to the reducibility of $U^+$ or $U^-$:
\vskip.5pc\noindent
{\bf C.3.u.}\quad ${\mathcal I}_{\mathcal P}(S_\mu^+,s)=
{\mathcal I}_{\mathcal P}(S_\mu^+,s,U^+{\rm irred.})\cup {\mathcal I}_{\mathcal P}(S_\mu^+,s,U^+{\rm red.})$,
\vskip.5pc\noindent
{\bf C.3.d.}\quad ${\mathcal I}_{\mathcal P}(S_\mu^-,s)=
{\mathcal I}_{\mathcal P}(S_\mu^-,s,U^-{\rm irred.})\cup {\mathcal I}_{\mathcal P}(S_\mu^-,s,U^-{\rm red.})$,
\vskip.5pc\noindent
{\bf C.3.s.}\quad ${\mathcal I}_{\mathcal P}(S_\mu^+\cup S_\mu^-,s)=
{\mathcal I}_{\mathcal P}(S_\mu^+\cup S_\mu^-,s,U^\pm{\rm irred.})\cup {\mathcal I}_{\mathcal P}(S_\mu^+\cup S_\mu^-,s,U^+\;{\rm or}\; U^-\,{\rm red.})$,
\vskip.5pc\noindent
with obvious meaning of the notation.
\vskip.8pc\noindent
In section 4 we completely identify the possible irreducible generalized projective representations $U$ of $\mathcal P$ for which
$U^+$ and $U^-$ are {\sl irreducible},
i.e. the components ${\mathcal I}_{\mathcal P}(S_\mu^+,s,U^+{\rm irred.})$, ${\mathcal I}_{\mathcal P}(S_\mu^-,s,U^-{\rm irred.})$
and ${\mathcal I}_{\mathcal P}(S_\mu^+\cup S_\mu^-,s,U^\pm{\rm irred.})$ of the decompositions {\bf C.3.u.}, {\bf C.3.d.} and {\bf C.3.s.}.
In doing so we shall identify, besides the well known irreducible generalized projective representations $U$ of $\mathcal P$
with $\TREV$ anti-unitary and $\SREV$ unitary, or with $\TREV$ unitary and $\SREV$ unitary, also those with $\TREV$ anti-unitary and $\SREV$ anti-unitary,
or with $\TREV$ unitary and $\SREV$ anti-unitary, which are not taken into account in the literature about elementary particles theory.
The class ${\mathcal I}_{\mathcal P}(S_\mu^+\cup S_\mu^-,s,U^+\;{\rm or}\; U^-\,{\rm red.})$ is investigated in section 5.
\section{Irreducible $U$ with $U^\pm$ irreducible}
\subsection{The case $\sigma(\underline P)=S_\mu^\pm$ with $U^\pm$ irreducible}
The irreducible generalized  projective representation of ${\mathcal P}$ with $\sigma(\underline P)=S_\mu^+$ (resp., $\sigma(\underline P)=S_\mu^+$) and
with $U^+$ (resp., $U^-$) irreducible are well known \cite{W2},\cite{c13}.
For each allowed pair $\mu>0$ and $s\in\frac{1}{2}\NN$ of the parameters characterizing
the representation,
{\sl modulo unitary isomorphisms} there is only one irreducible projective representation
of ${\mathcal P}_+^\uparrow$ with $\sigma(\underline P)=S_\mu^+$ and only one with $\sigma(\underline P)=S_\mu^-$, that we report.
The Hilbert space of the projective representation is the space $L_2(\RR^3,\CC^{2s+1},d\nu)$ of all functions
$\psi:\RR^3\to \CC^{2s+1}$,
${\bf p}\to\psi({\bf p})$, square integrable with respect to the invariant measure
$d\nu({\bf p})=\frac{dp_1dp_2dp_3}{\sqrt{\mu^2+{\bf p}^2}}$.
The irreducible generalized representations of $\mathcal P$ are obtained by adding $\TREV$ and $\SREV$, accorind the next sections 4.1.2 and 4.1.3.
\subsubsection{The case $\sigma(\underline P)=S_\mu^+$}
Fixed $\mu$ and $s$, for the irreducible generalized projective representation with $\sigma(\underline P)=S_\mu^+$ the following statements hold.
\begin{itemize}
\item[--]
The generators $P_j$ are the multiplication operators defined by
$(P_j\psi)({\bf p})=p_j\psi({\bf p})$; as consequence
\item[--]
$(P_0\psi)({\bf p})=p_0\psi({\bf p})$ where $p_0=+\sqrt{\mu^2+{\bf p}^2}$,
because $P_0$ has a positive spectrum;
\item[--]
the generators $J_j$ are given by $J_j=i\left(p_k\frac{\partial}{\partial p_l}-p_l\frac{\partial}{\partial p_k}\right)+S_j$,
$(j,k,l)$ being a cyclic permutation of $(1,2,3)$,
where $S_1,S_2,S_3$ are the self-adjoint generators of an irreducible projective representation $L:SO(3)\to\CC^{2s+1}$
such that $S_1^2+S_2^2+S_3^2=s(s+1)\Id$;
hence, they can be fixed to be the three spin operators of $\;\CC^{2s+1}$;
\item[--]
the generators $K_j$ are given by
$K_j=ip_0\frac{\partial}{\partial p_j}-\frac{({\bf S}\land {\bf p})_j}{\mu+p_0}$;
\item[--]
the space inversion -- unitary  -- and the time reversal -- anti-unitary -- operators are
$$
\SREV=\Upsilon,\quad\hbox{and}\quad\TREV=\tau{\mathcal K}\Upsilon\,,\quad\hbox{where}
\eqno(19)
$$
\item[]
-\quad$\Upsilon$ is the unitary operator defined by
$(\Upsilon\psi)({\bf p})=\psi(-{\bf p})$,
\item[]
-\quad
$\tau$ is a unitary matrix of \,$\CC^{2s+1}$ such that $\tau {\overline S_j}\tau^{-1}= -{S_j}$, for all $j$;
such a matrix always exists and it is unique up a complex factor of modulus 1;
moreover,
if $s\in\NN$ then $\tau$ is symmetric and $\tau\overline\tau=1$, while if $s\in\left(\NN+\frac{1}{2}\right)$ then $\tau$ is
anti-symmetric and $\tau\overline\tau=-1$ \cite{c13};
\item[]
-\quad
$\mathcal K$ is the anti-unitary complex conjugation operator defined by ${\mathcal K}\psi({\bf p})=\overline{\psi({\bf p})}$.
\end{itemize}
\subsubsection{The case $\sigma(\underline P)=S_\mu^-$.}
For the irreducible projective representation with characterizing parameters $\mu$, $s$ and
$\sigma(\underline P)=S_\mu^-$, the following {\sl symmetrical}
statements hold.
\begin{itemize}
\item[--]
The generators $P_j$ are the multiplication operators defined by
$(P_j\psi)({\bf p})=p_j\psi({\bf p})$; as consequence
\item[--]
$(P_0\psi)({\bf p})=-p_0\psi({\bf p})$,
because $P_0$ has a negative spectrum;
\item[--]
the generators $J_j$ are given by  $J_j=i\left(p_k\frac{\partial}{\partial p_l}-p_l\frac{\partial}{\partial p_k}\right)+S_j$,
$(j,k,l)$ being a cyclic permutation of $(1,2,3)$;
\item[--]
the generators $K_j$ are given by
$K_j=-ip_0\frac{\partial}{\partial p_j}+\frac{({\bf S}\land {\bf P})_j}{\mu+p_0}$;
\item[--]
the space inversion and time reversal operators are
$\SREV=\Upsilon$ and $\TREV=\tau{\mathcal K}\Upsilon$.
\end{itemize}
\subsection{The case $\sigma(\underline P)=S_\mu^+\cup S_\mu^-$ with $U^+$ irreducible}
Now we establish results that allow us to identify all the irreducible generalized projective representations
with $\sigma (\underline P)=S_\mu^+\cup S_\mu^-$ and $U^\pm$ irreducible.
Prop.s 3.4-3.6 imply that
$\TREV$ is unitary or $\SREV$ is anti-unitary. Moreover, according to Prop. 3.3, $U\mid_{{\mathcal P}_+^\uparrow}$ is always reduced by
$E^+=E(S_\mu^+)\equiv E^{(0)}[\mu,\infty)=\chi_{[\mu,\infty)}(P_0)$ and
$E^-=E(S_\mu^-)\equiv E^{(0)}(-\infty,-\mu]=\chi_{(-\infty,-\mu]}(P_0)$, so that for all $g\in{\mathcal P}_+^\uparrow$
we can write $U_g=U^+_g+U^-_g$, where $U^+_g=E^+U_gE^+$ and $U^-_g=E^-U_gE^-$.
\par
Each of these two components $U^+$ and
$U^-$ can be reducible or not, in its turn.
The following proposition entails that the reducibility of $U^+$ is equivalent to the reducibility of
$U^-$.
\vskip.5pc\noindent
{\bf Proposition 4.1.}
{\sl
Let $U$ belong to ${\mathcal I}_{\mathcal P}(S_\mu^+\cup S_\mu^-$.
If $F_+$ is a projection operator that reduces $U^+$, then the following statements hold.
\begin{itemize}
\item[(i)]
In the case that $\TREV$ is unitary, the projection operator $F_-^\trev=\TREV F_+\TREV$ reduces
$U^-$, and $F^\trev=F_++ F_-^\trev$ reduces $U\mid_{{\mathcal P}_+}$;
\item[(ii)]
in the case that $\SREV$ is anti-unitary, the projection operator $F_-^\srev=\SREV F_+\SREV$ reduces
$U^-$, and $F^\srev=F_++ F_-^\srev$ reduces $U\mid_{{\mathcal P}^\uparrow}$.
\end{itemize}}
\vskip.4pc\noindent
{\bf Proof.}
If $\TREV$ is unitary, then
$\TREV^{-1}=\TREV$ and $\TREV P_0\TREV=-P_0$ follow from (7); this implies
$\TREV E^+\TREV=\TREV\chi_{[\mu,\infty)}(P_0)\TREV=\chi_{(-\infty,-\mu]}=E^-$ by Lemma 3.1.
If $F_+$ is a projection operator that reduces $U^+$, and hence $\nop<F_+< E^+$,
then
$F_-^\trev E^-=(\TREV F_+\TREV)E^-=(\TREV F_+\TREV)\TREV E^+\TREV=\TREV F_+ E^+\TREV=\TREV F_+ \TREV$ since $F_+< E^+$.
Therefore, $\nop< F_-^\trev< E_-$ is satisfied.
Now we show that
$[F_-^\trev,P_0^-]=[F_-^\trev,P_j^-]=[F_-^\trev,K_j^-]=[F_-^\trev,J_k^-]=\nop$, i.e. that
$F_-^\trev$ reduces $U^-$.
Since $P_0^-=E^-P_0E^-$ and $[F_+,P_0]=[F_+,P_0^+]=\nop$,
we have
$$\begin{array}{ll}
P_0^-F_-^\trev&=P_0^-F_-^\trev E^-=
E^-P_0E^-\TREV F_+\TREV E^-=
E^-P_0\TREV E^+\TREV \TREV F_+\TREV E^-\cr
&=-E^-\TREV P_0 E^+ F_+ \TREV E^-=
-E^-\TREV E^+ P_0 F_+ \TREV E^-=
-E^-\TREV E^+F_+ P_0\TREV E^-\cr
&=-E^-\TREV  F_+ P_0\TREV E^-=
E^-\TREV F_+\TREV  P_0 E^-=
E^-F_-^\trev P_0 E^-=
F_-^\trev E^- P_0E^-=
F_-^\trev P_0^-.
\end{array}
$$
A similar derivation shows that
$[F_-^\trev,P_j]=[F_-^\trev,K_j^-]=[F_-^\trev,J_k^-]=\nop$; therefore $F_-^\trev$ reduces $U^-$.
Now we see that $F^\trev=F_++F_-^\trev$ reduces $U\mid_{{\mathcal P}_+}$. The equalities
$F^\trev P_0=(F_++F_-^\trev)P_0=P_0(F_++F_-^\trev)=P_0F^\trev$ immediately follow from
$P_0=E^+P_0 E^++E^-P_0E^-$ and $F_-^\trev E^-=F_-^\trev$, $F_+E^+=F_+$, $F_+E^-=F_-^\trev E^+=\nop$;
similarly,
$[F_-^\trev,P_j]=[F_-^\trev,J_k]=[F_-^\trev,K_j]=\nop$ hold. Hence, $F^\trev$ reduces $U\mid_{{\mathcal P}_+^\uparrow}$.
Moreover, $F^\trev\TREV=F_+\TREV+F_-^\trev\TREV=\TREV\TREV F_+\TREV+\TREV F_+\TREV\TREV=\TREV F_-^\trev+\TREV F_+=\TREV F^\trev$.
Therefore, $F^\trev$ reduces also $U\mid_{{\mathcal P}_+}$.
A quite similar argument proves statement (ii).
\hfill{$\bullet$}
\vskip.5pc\noindent
{\bf Corollary.}
{\sl Under the hypotheses of Prop. 3.1, $U^+$ is reducible if and only if  $U^-$ is reducible.}
\vskip.5pc\noindent
Prop. 3.1 and its corollary indicate that the irreducible generalized projective representations of $\mathcal P$
can be classified according to the reducibility of $U^+$.
\subsubsection{Hilbert space and self-adjoint generators.}
In the case that $U^+$ is irreducible, with $\sigma(\underline P)=S_\mu^+\cup S_\mu^-$,
according to Prop. 3.3 the restriction $U:{\mathcal P}_+^\uparrow\to{\mathcal U}({\mathcal H})$ must be the direct sum of
$U^+:{\mathcal P}_+^\uparrow\to{\mathcal U}({\mathcal H}^+)$ and $U^-:{\mathcal P}_+^\uparrow\to{\mathcal U}({\mathcal H}^-)$,
where ${\mathcal H}^+=E^+({\mathcal H})$, ${\mathcal H}^-=E^-({\mathcal H})$ and ${\mathcal H}^+\oplus{\mathcal H}^-={\mathcal H}$;
according to Prop. 4.1, both $U^+$ and $U^-$
are irreducible projective representations of ${\mathcal P}_+^\uparrow$. Since $\sigma(P_0^+)=[\mu,\infty)$
(resp., $\sigma(P_0^-)=(-\infty,-\mu]$), the reduced projective representation $U^+$ (resp., $U^-$) is
unitarily isomorphic to the projective representation $U:{\mathcal P}_+^\uparrow\to {\mathcal U}(L_2(\RR^3,\CC^{2s+1},d\nu))$ with
$\sigma(\underline P)=S_\mu^+$
(resp., with $\sigma(\underline P)=S_\mu^-$) described in sect. 4.1, with the same characterizing parameters $\mu$ and $s$ of the
unrestricted irreducible generalized projective representation $U$.
Accordingly, there are two unitary mappings $W^+:{\mathcal H}^+\to L_2(\RR^3,\CC^{2s+1},d\nu)$, and $W^-:{\mathcal H}^-\to L_2(\RR^3,\CC^{2s+1},d\nu)$
such that the reduced generators in the Hilbert space $W^+({\mathcal H}^+)\equiv L_2(\RR^3,\CC^{2s+1},d\nu)$ are the following.
\par
$P_0^+=W^+(E^+P_0E^+)$, so that
$(P_0^+\phi)({\bf p})=p_0({\bf p})\phi({\bf p})=\sqrt{\mu^2+{\bf p}^2}\phi({\bf p})$;
\par
$P_j^+=p_j$;
\par
$J_k^+={\textsf j}_k=i\left(p_l\frac{\partial}{\partial p_j}-p_j\frac{\partial}{\partial p_l}\right)+S_k$;
\par
$K_j^+={\textsf k}_j=ip_0\frac{\partial}{\partial p_j}-\frac{({\bf S}\land {\bf P})_j}{\mu+p_0}$.
\par\noindent
Symmetrically,
the reduced generators in $W^-({\mathcal H}^-)\equiv L_2(\RR^3,\CC^{2s+1},d\nu)$ are
\par\noindent
$P_0^-=W^-(E^-P_0E^-)=-p_0$; $P_j^-=p_j$;
$J_k^-={\textsf j}_k=i\left(p_l\frac{\partial}{\partial p_j}-p_j\frac{\partial}{\partial p_l}\right)+S_k$;
$K_j^-=-{\textsf k}_j=-ip_0\frac{\partial}{\partial p_j}+\frac{({\bf S}\land {\bf p})_j}{\mu+p_0}$.
\par
Hence we have proved that, modulo unitary isomorphisms, the Hilbert space of the representation
is $L_2(\RR^3,\CC^{2s+1},d\nu)\oplus L_2(\RR^3,\CC^{2s+1},d\nu)$.
It is convenient to represent each vector $\psi\in{\mathcal H}=E^+\psi+E^-\psi$ as a column vector
$\psi=\left[\begin{array}{c}\psi^+\cr \psi^-\end{array}\right]$, where $\psi^+=W^+(E^+\psi)$ and $\psi^-=W^-(E^-\psi)$;
in such a representation the generators of $U\mid_{{\mathcal P}_+^\uparrow}$ take
the following form, known as the {\sl canonical form}.
$$
P_j=\left[\begin{array}{cc}p_j&0\cr 0&-p_j\end{array}\right],\;\;
P_0=\left[\begin{array}{cc}p_0&0\cr 0&-p_0\end{array}\right],\;\;
J_k=\left[\begin{array}{cc}{\textsf j}_k&0\cr 0&{\textsf j}_k\end{array}\right],\;\;
K_j=\left[\begin{array}{cc}{\textsf k}_j&0\cr 0&-{\textsf k}_j\end{array}\right].\eqno(20)
$$
\subsubsection{Time reversal and space inversion operators.}
The condition $\sigma(\underline P)=S_\mu^+\cup S_\mu^-$ implies that
the time reversal operator $\TREV$ must be unitary or the space inversion operator $\SREV$
must be anti-unitary, according to Prop.s 3.4-3.6.
In the case in which both $\TREV$ {\sl and} $\SREV$ are
unitary their explicit form is well known, up a complex factor of modulus 1 \cite{c13}.
$$
\TREV=\left[\begin{array}{cc}0&1\cr 1&0\end{array}\right]\,;\qquad \SREV=\Upsilon\left[\begin{array}{cc}1&0\cr 0&1\end{array}\right]
\quad\hbox{or}\quad\SREV=\Upsilon\left[\begin{array}{cc}1&0\cr 0&-1\end{array}\right]\,.
\eqno(21)
$$
(In the matrices (21) ``$1$'' and ``$0$'' denote the identity and null operators of\;  $\CC^{2s+1}$. This notation is adopted throughout the paper,
whenever it does not cause confusions)
\vskip.5pc\noindent
However, irreducible generalized projective representations with $\TREV$ anti-unitary or $\SREV$ anti-unitary do exist,
as we show after the following Prop.4.2.
\vskip.5pc\noindent
{\bf Proposition 4.2.}
{\sl
Let $U:{\mathcal P}\to{\mathcal V}({\mathcal H})$ be an irreducible generalized projective representation of $\mathcal P$,
with $U^+$ irreducible. The following statements hold.
\begin{itemize}
\item[i)]
If $\TREV$ is anti-unitary then
\item[]
$\TREV=\tau{\mathcal K}\Upsilon\left[\begin{array}{cc}1&0\cr 0&e^{i\theta}\end{array}\right]$;
hence, $\TREV$ can be taken to be $\TREV=\tau{\mathcal K}\Upsilon\left[\begin{array}{cc}1&0\cr 0&1\end{array}\right]$
up a complex factor of modulus 1;
\item[] in particular,
$s\in\NN$ implies $\TREV^2=\Id$\quad and\quad $s\in(\NN+\frac{1}{2})$ implies $\TREV^2=-\Id$;
\item[ii)]
if $\SREV$ is anti-unitary then \item[]
$\SREV=\left[\begin{array}{cc}0&\tau\cr \tau&0\end{array}\right]{\mathcal K}$ when
$\SREV^2=\Id$ and $s\in\NN$, or when $\SREV^2=-\Id$ and $s\in(\NN+\frac{1}{2})$,
\item[]
$\SREV=\left[\begin{array}{cc}0&\tau\cr -\tau&0\end{array}\right]{\mathcal K}$ when
$\SREV^2=-\Id$ and $s\in\NN$, or when $\SREV^2=\Id$ and $s\in(\NN+\frac{1}{2})$.
\end{itemize}}
\noindent
{\bf Proof.}
Since $\TREV$ is anti-unitary,
the operator $\hat T=\tau{\mathcal K}\Upsilon\TREV\equiv\left[\begin{array}{cc} T_{11}&T_{12}\cr T_{21}&T_{22}\end{array}\right]$
is unitary, and $\TREV={\mathcal K}\Upsilon\tau^{-1}\hat T$. Now, (5), $\tau^{-1}\overline{S_k}\tau=-S_k$ and the properties
\par\noindent
$\Upsilon p_j=-p_j\Upsilon$, $\Upsilon\frac{\partial}{\partial p_j}=-\frac{\partial}{\partial p_j}\Upsilon$, $\Upsilon^2=\Id$,
${\mathcal K}p_j=p_j{\mathcal K}$, ${\mathcal K}\frac{\partial}{\partial p_j}=\frac{\partial}{\partial p_j}{\mathcal K}$,
${\mathcal K}\Upsilon=\Upsilon{\mathcal K}$, ${\mathcal K}^2=\Id$
\par\noindent
imply $[\hat T,P_j]=\nop$, $[\hat T,P_0]=\nop$, $[\hat T,J_k]=\nop$. The first two of these last three equalities imply that
$\hat T=\left[\begin{array}{cc} T_1({\bf p})&0\cr 0& T_2({\bf p})\end{array}\right]$, where
$T_m({\bf p})$ is a $(2s+1)\times(2s+1)$ matrix for every ${\bf p}\in\RR^3$, so that $[T_m({\bf p}),p_j]=\nop$; the third equality implies
$[T_m({\bf p}),{\textsf j}_k]=\nop$.
Then, since $p_1,p_2,p_3,{\textsf j}_1,{\textsf j}_2,{\textsf j}_3$
are the generators of an irreducible projective representation of $\mathcal E$
in the Hilbert space $L_2(\RR^3,\CC^{2s+1},d\nu)$, the relations $[T_m({\bf p}),p_j]=\nop$ and
$[T_m({\bf p}),{\textsf j}_k]=\nop$ imply
$T_m({\bf p})=e^{i\theta_m}\Id$, i.e.
$\hat T=\left[\begin{array}{cc} e^{i\theta_1}&0\cr 0& e^{i\theta_2}2\end{array}\right]$, with $\theta_{1,2}$ constant. Hence we have
$\TREV={\mathcal K}\Upsilon\tau^{-1}\left[\begin{array}{cc} 1&0\cr 0& e^{i\theta}\end{array}\right]=\pm\tau{\mathcal K}\Upsilon\left[\begin{array}{cc} e^{i\theta_1}&0\cr 0& e^{i\theta_2}\end{array}\right]$; the free phase can be chosen so that
$\TREV=\tau{\mathcal K}\Upsilon\left[\begin{array}{cc} 1&0\cr 0& e^{i\theta}\end{array}\right]$.
By transforming each operator $B$ into $WBW^{-1}$, where $W=\left[\begin{array}{cc} 1&0\cr 0& e^{i\frac{\theta}{2}}\end{array}\right]$,
$\TREV$ turns out to be transformed into $\TREV=\tau{\mathcal K}\Upsilon\left[\begin{array}{cc} 1&0\cr 0& 1\end{array}\right]$, while all generators
$P_j$, $P_0$, $J_k$, $K_j$ remain unchanged.
Accordingly, $\TREV^2=\left[\begin{array}{cc} \tau\overline\tau&0\cr 0& \tau\overline\tau\end{array}\right]$.
If $s\in\NN$, then $\tau\overline\tau=1$, so that $\TREV^2=\Id$; if $s\in(\NN+\frac{1}{2})$, then $\tau\overline\tau=-1$, so that
$\TREV^2=-\Id$. This proves (i).
\par
The proof of (ii) is carried out along quite similar lines.
\hfill{$\bullet$}
\vskip.5pc\noindent
In an irreducible generalized projective representation of $\mathcal P$ with $\sigma(\underline P)=S_\mu^+\cup S_\mu^-$
the combination $\TREV$ anti-unitary and $\SREV$ unitary is excluded by Prop.3.4.
However, all other combinations can occur. The combination $\TREV$ unitary and $\SREV$ unitary
is already settled according to (21).
Then we check the consistency of the remaining combinations;
it is sufficient to verify that (6) is satisfied, since all other
conditions (1)-(5) for $P_J$, $P_0$, $J_k$, $K_j$, $\TREV$ and $\SREV$ are valid by construction.
\begin{itemize}
\item[a)]
If {\sl $\TREV$ is anti-unitary and $\SREV$ is anti-unitary}, then they have the form shown by Prop. 4.2.
By a straightforward calculation it is verified that condition (6) is always satisfied.
\item[b)]
If {\sl $\TREV$ is unitary and $\SREV$ is anti-unitary}, then they have the form given by (21) and (ii) in Prop. 4.2.
We see that (6) is always satisfied.
\end{itemize}
Thus, besides the usually considered irreducible generalized projective representations with the combination $\TREV$ unitary, $\SREV$ unitary,
also the combinations $\TREV$ unitary, $\SREV$  anti-unitary and $\TREV$ anti-unitary, $\SREV$  anti-unitary can occur.
They are completely determined according to the following scheme.
\subsection{Resulting scheme}
For every $\mu>0$ and every $s\in\frac{1}{2}\NN$ there are height irreducible generalized projective representations of $\mathcal P$:
\begin{itemize}
\item[{\bf u.}]
the representation with {\sl up} spectrum $\sigma(\underline P)=S_\mu^+$, identified in section 4.1.1;
\item[{\bf d.}]
the representation with {\sl down} spectrum $\sigma(\underline P)=S_\mu^-$, identified in section 4.1.2;
\item[{\bf s.}]
six inequivalent representations $U^{(1)}$, $U^{(2)}$,...,$U^{(6)}$ with {\sl symmetrical} spectrum $\sigma(\underline P)=S_\mu^+\cup S_\mu^-$,
identified in section 4.2, all with the same Hilbert space ${\mathcal H}=L_2(\RR^3,\CC^{2s+1},d\nu)\oplus L_2(\RR^3,\CC^{2s+1},d\nu)$
and the same self-adjoint generators (20); they differ just for the different combinations of time reversal and space inversion operators.
\item[] $U^{(1)}$ has unitary $\TREV=\left[\begin{array}{cc}0&1\cr 1&0\end{array}\right]$ and unitary $\SREV=\Upsilon\left[\begin{array}{cc}1&0\cr 0&1\end{array}\right]$;
\item[] $U^{(2)}$ has unitary $\TREV=\left[\begin{array}{cc}0&1\cr 1&0\end{array}\right]$ and unitary $\SREV=\Upsilon\left[\begin{array}{cc}1&0\cr 0&-1\end{array}\right]$;
\item[] $U^{(3)}$ has unitary $\TREV=\left[\begin{array}{cc}0&1\cr 1&0\end{array}\right]$ and
anti-unitary $\SREV=\left[\begin{array}{cc}0&\tau\cr \tau&0\end{array}\right]{\mathcal K}$;
\item[] $U^{(4)}$ has unitary $\TREV=\left[\begin{array}{cc}0&1\cr 1&0\end{array}\right]$ and
anti-unitary $\SREV=\left[\begin{array}{cc}0&\tau\cr -\tau&0\end{array}\right]{\mathcal K}$;
\item[] $U^{(5)}$ has anti-unitary $\TREV=\tau{\mathcal K}\Upsilon\left[\begin{array}{cc}0&1\cr 1&0\end{array}\right]$ and anti-unitary$\SREV=\left[\begin{array}{cc}0&\tau\cr\tau&0\end{array}\right]{\mathcal K}$;
\item[] $U^{(6)}$ has anti-unitary $\TREV=\tau{\mathcal K}\Upsilon\left[\begin{array}{cc}0&1\cr 1&0\end{array}\right]$ and anti-unitary$\SREV=\left[\begin{array}{cc}0&\tau\cr-\tau&0\end{array}\right]{\mathcal K}$.
\end{itemize}
The class of all such octets, for all permitted values of $\mu$ and $s$, does not exhaust ${\mathcal I}_{\mathcal P}$, because the components with $U^+$ or  $U^-$ reducible in the decompositions {\bf C.3} are not empty, as we show in section 5.
\section{Irreducible $U:{\mathcal P}\to{\mathcal V}({\mathcal H})$ with $U^+$ or $U^-$ reducible}
The current relativistic quantum theories of a particle are developed only on irreducible generalized projective representations
$U:{\mathcal P}\to{\mathcal V}({\mathcal H})$ with $U^+$ and $U^-$ irreducible.
This would be a correct practice
if the irreducibility of the whole $U$ implied the irreducibility of the reductions
$U^\pm=E^\pm U\mid_{{\mathcal P}_+^\uparrow}E^\pm$.
This is not the case.
In this section, in fact, we show that irreducible generalized projective representations $U$ of $\mathcal P$ exist such that
$U^\pm$ is reducible in the case $\sigma(\underline P)=S_\mu^\pm$, as well as in the case $\sigma(\underline P)=S_\mu^+\cup S_\mu^-$.
\subsection{The cases $\sigma(\underline P)=S_\mu^\pm$}
Given an irreducible generalized projective representation of $\mathcal P$,
Prop. 3.3 implies that if the restriction
$U\mid_{{\mathcal P}_+^\uparrow}$ is irreducible too, then either
$\sigma(\underline P)=S_\mu^+$ or $\sigma(\underline P)=S_\mu^-$.
The converse is not true; in other words, the condition $\sigma(\underline P)=S_\mu^+$
implies $U\mid_{{\mathcal P}_+^\uparrow}=U^+$, but does not imply
that $U^+$ is irreducible.
In fact, now we identify irreducible generalized projective representations $U:{\mathcal P}\to{\mathcal V}({\mathcal H})$
for which $U^+$ is reducible.
\par
We deal with the case $\sigma(\underline P)=S_\mu^+$;
the alternative  case $\sigma(\underline P)=S_\mu^-$ can be addressed along identical lines.
We show that for any $\mu>0$ and any $s\in\frac{1}{2}\NN$ there are irreducible generalized projective
representations $U$ of $\mathcal P$ such that $U^+$ is the direct sum $U^{(1)}\oplus U^{(2)}$ of two identical projective representations
$U^{(1)}:{\mathcal P}_+^\uparrow\to {\mathcal U}({\mathcal H}^{(1)})$ and
$U^{(2)}:{\mathcal P}_+^\uparrow\to {\mathcal U}({\mathcal H}^{(2)})$.
\par
Let us consider two irreducible projective representations
$U^{(1)}:{\mathcal P}_+^\uparrow\to {\mathcal U}({\mathcal H}^{(1)})$ and
$U^{(2)}:{\mathcal P}_+^\uparrow\to {\mathcal U}({\mathcal H}^{(2)})$ of ${\mathcal P}_+^\uparrow$ of the form
described in sect. 4.1.1, with the same pair $\mu$, $s$ of parameters that determine the representations up unitary isomorphisms,
and with ${\mathcal H}^{(1)}={\mathcal H}^{(2)}=L_2(\RR^3,\CC^{2s+1},d\nu)$.
The Hilbert space of the direct sum $U^{(1)}\oplus U^{(2)}$ is
${\mathcal H}= L_2(\RR^3,\CC^{2s+1},d\nu) \oplus  L_2(\RR^3,\CC^{2s+1},d\nu)$.
\par
It is convenient to represent
every vector $\psi=\psi_1+\psi_2$ in $\mathcal H$, with $\psi_1\in{\mathcal H}^{(1)}$ and $\psi_2\in{\mathcal H}^{(2)}$,
as the column vector $\psi\equiv \left[\begin{array}{c}\psi_1\cr \psi_2\end{array}\right]$, so that every linear (resp., anti-linear) operator $A$ of $\mathcal H$
can be represented by a matrix $\left[\begin{array}{cc}A_{11}& A_{12} \cr A_{21} & A_{22}\end{array}\right]$, where $A_{mn}$ is a linear (resp., anti-linear)
operator of $L_2(\RR^3,\CC^{2s+1},d\nu)$, and
$A\psi=\left[\begin{array}{cc}A_{11} & A_{12} \cr A_{21} & A_{22}\end{array}\right]  \left[\begin{array}{c}\psi_1\cr \psi_2\end{array}\right]=
\left[\begin{array}{c}
A_{11}\psi_1+A_{12}\psi_2 \cr A_{21}\psi_1+A_{22}\psi_2\end{array}\right]$.
Let us introduce the following operators of $\mathcal H$.
$$
P_0=\left[\begin{array}{cc}p_0& 0 \cr 0 & p_0\end{array}\right]\,,\quad
P_j=\left[\begin{array}{cc}p_j& 0 \cr 0 & p_j\end{array}\right]\,,\quad
J_k=\left[\begin{array}{cc}{\textsf j}_k& 0 \cr 0 & {\textsf j}_k\end{array}\right]\,,\quad
K_j=\left[\begin{array}{cc}{\textsf k}_j& 0 \cr 0 & {\textsf k}_j\end{array}\right]\,,\eqno(22)
$$
where
${\textsf j}_k=i\left(p_l\frac{\partial}{\partial p_j}-p_j\frac{\partial}{\partial p_l}\right)+S_k$ and
${\textsf k}_j=ip_0\frac{\partial}{\partial p_j}-\frac{({\bf S}\land {\bf p})_j}{\mu+p_0}$.
\par
These operators are self-adjoint and satisfy relations (1). Then (22) are the generators of a
reducible projective representation $U:{\mathcal P}_+^\uparrow\to L_2(\RR^3,\CC^{2s+1},d\nu)\oplus L_2(\RR^3,\CC^{2s+1},d\nu)$.
Since for this representation $\sigma(\underline P)=S_\mu^+$,
the possible extensions to the whole $\mathcal P$ are obtained by introducing a time reversal operator $\TREV$ and a space inversion operator $\SREV$
in such a way to satisfy (2), (5), (6).
In sections 5.1.1 and 5.1.2 we show that, while fixed $\mu$ and $s$ there is a unique such possibility for $\SREV$ up to unitary equivalence,
there are inequivalent possibilities for $\TREV$. In section 5.1.3 we prove that some of these possibilities give rise to irreducible
generalized projective representations of of $\mathcal P$.
\subsubsection{Space inversion operator $\SREV$.}
According to Prop. 3.4-3.6,
the condition $\sigma(\underline P)=S_\mu^+$ implies that $\TREV=\left[\begin{array}{cc}\TREV_{11}& \TREV_{12} \cr \TREV_{21} & \TREV_{22}\end{array}\right]$ is anti-unitary
and $\SREV = \left[\begin{array}{cc}\SREV_{11}&\SREV_{12} \cr\SREV_{21} &\SREV_{22}\end{array}\right]$ is unitary.
We begin by determining $\SREV$. Relations (2) imply
$$
\SREV_{mn}p_0=p_0\SREV_{mn},\; \SREV_{mn}p_j=-p_j\SREV_{mn},\; \SREV_{mn}{\textsf j}_k={\textsf j}_k\SREV_{mn},\; \SREV_{mn}{\textsf k}_j=-{\textsf k}_j\SREV_{mn}.\eqno(23)
$$
The unitary operator
$\Upsilon$ defined on $\mathcal H$ satisfies the following relations.
$$
\Upsilon p_j=-p_j\Upsilon,\quad\Upsilon\frac{\partial}{\partial p_j}=-\frac{\partial}{\partial p_j}\Upsilon,\quad[\Upsilon, S_j]=\nop,\quad \Upsilon^2=\Id\,.
\eqno(24)
$$
Once introduced the unitary operator $\hat S=\Upsilon\SREV$, from (2), (24) and (23) we derive
\vskip.8pc\noindent
${\hat S}_{mn}p_0\equiv({\hat S}P_0)_{mn}=\Upsilon\SREV_{mn}p_0=p_0\Upsilon\SREV_{mn}=p_0\hat S_{mn}$,\hfill{(25.i)}
\vskip.7pc\noindent
${\hat S}_{mn}p_j\equiv({\hat S}P_j)_{mn}=\Upsilon\SREV_{mn}p_j=-\Upsilon p_j\SREV_{mn}=p_j\Upsilon\SREV_{mn}=p_j\hat S_{mn}$,\hfill{(25.ii)}
\vskip.7pc\noindent
${\hat S}_{mn}{\textsf j}_k\equiv({\hat S}J_k)_{mn}=\Upsilon\SREV_{mn}{\textsf j}_k=\Upsilon {\textsf j}_k\SREV_{mn}=
\Upsilon\left(ip_l\frac{\partial}{\partial p_j} -ip_j\frac{\partial}{\partial p_l}+S_k\right)\SREV_{mn}=$\hfill{(25.iii)}
\vskip.5pc
$\qquad\quad=\left(ip_l\frac{\partial}{\partial p_j}-ip_j\frac{\partial}{\partial p_l}+S_k\right)\Upsilon\SREV_{mn}={\textsf j}_k{\hat S}_{mn}$,
\vskip.7pc\noindent
${\hat S}_{mn}k_j\equiv({\hat S}K_j)_{mn}=\Upsilon\SREV_{mn}k_j=-\Upsilon k_j\SREV_{mn}=
-\Upsilon\left(ip_0\frac{\partial}{\partial p_j}-\frac{[{\bf S}\land{\bf p}]_j}{\mu+p_0}\right)\SREV_{mn}=$\hfill{(25.iv)}
\vskip.5pc
$\qquad\quad=\left(ip_0\frac{\partial}{\partial p_j}-\frac{[{\bf S}\land{\bf p}]_j}{\mu+p_0}\right)\Upsilon\SREV_{mn}=
{\textsf k}_j{\hat S}_{mn}$.
\vskip.8pc
Now, since each component projective representation $U^{(m)}:{\mathcal P}_+^\uparrow\to{\mathcal H}^{(m)})$
is irreducible, (25) imply that each ${\hat S}_{mn}$ is a multiple of the identity, so that
${\hat S}=\left[\begin{array}{cc}c_{11}& c_{12} \cr c_{21} & c_{22}\end{array}\right]$.
According to (2), the further constraint $\SREV^2=\Id$ can be imposed; it is satisfied if and only if ${\hat S}^2=\Id$;
this implies that $\hat S={\hat S}^{-1}={\hat S}^\ast$ is a constant hermitean matrix with eigenvalues
$+1$ and $-1$, where ${\hat S}^\ast$ denotes the adjoint of $\hat S$; therefore, a unit vector ${\bf n}\in\RR^3$ exists such that
$$
\SREV=\Upsilon\,{\bf n}\cdot{\vec\sigma},\;\hbox{ where }{\vec\sigma} =\left(\left[\begin{array}{cc}0&1\cr 1&0\end{array}\right],\left[\begin{array}{cc}0&-i\cr i&0\end{array}\right],
\left[\begin{array}{cc}1&0\cr 0&-1\end{array}\right]\right).\eqno(26)
$$
If $\hat W=\left[\begin{array}{cc}w_{11}&w_{12}\cr w_{21}& w_{22}\end{array}\right]$ is any constant unitary $2\times 2$ matrix, then
$[\hat W,P_0]=[\hat W,P_j]=[\hat W,J_k]=[\hat W,K_j]=\nop$. Such a matrix always exists such that
$\hat W\SREV\hat W^{-1}=\Upsilon\hat W{\bf n}\cdot{\vec\sigma}\hat W^{-1}=
\Upsilon\left[\begin{array}{cc}0& 1\cr 1& 0\end{array}\right]$; therefore, by converting every operator $B$ into $\hat WB\hat W^{-1}$
we obtain a unitarily equivalent irreducible representation of $\mathcal P$.
In so doing the operators
$P_0$, $P_j$, $J_k$ and $K_j$ remain unaltered because each of them has the form $\left[\begin{array}{cc}A&0\cr 0&A\end{array}\right]$.
Thus, up to a unitary isomorphism, $\SREV$ satisfies (2), (6) if and only if
$$\SREV=\Upsilon\left[\begin{array}{cc}0&1\cr 1&0\end{array}\right]\,.
\eqno(27)
$$
\subsubsection{Time reversal operator $\TREV$.}
Now we identify the time reversal operator $\TREV$ that completes the generalized projective representation of $\mathcal P$ that
extends the reducible projective representation
$U=U^{(1)}\oplus U^{(2)}$ of ${\mathcal P}_+^\uparrow$ to $\mathcal P$.
The conditions (5) imply
$$
\TREV_{mn}p_0=p_0\TREV_{mn}\,,\; \TREV_{mn}p_j=-p_j\TREV_{mn}\,,\; \TREV_{mn}{\textsf j}_k=
-{\textsf j}_k\TREV_{mn}\,,\; \TREV_{mn}{\textsf k}_j={\textsf k}_j\TREV_{mn}.
\eqno(28)
$$
The anti-unitary operator ${\mathcal K}$
satisfies the following relation
$$
{\mathcal K}p_j=p_j{\mathcal K},\quad  {\mathcal K}\frac{\partial}{\partial p_j}=\frac{\partial}{\partial p_j}{\mathcal K},\quad
{\mathcal K}\Upsilon=\Upsilon{\mathcal K},\quad{\mathcal K}^2=\Id\,.\eqno(29)
$$
Let us introduce the operator $\hat T=\left[\begin{array}{cc}{\hat T}_{11}& {\hat T}_{12}\cr {\hat T}_{21}&{\hat T}_{22}\end{array}\right]$,
with ${\hat T}_{mn}=\tau{\mathcal K}\Upsilon\TREV_{mn}$, that is unitary,
so that
$\TREV=\Upsilon{\mathcal K}\tau^{-1}\hat T\equiv\tau{\mathcal K}\Upsilon\hat T$. Relations (28), (24), (29) imply
\vskip.8pc\noindent
$\hat T_{mn} p_0=\tau{\mathcal K}\Upsilon\TREV_{mn}p_0=
\tau{\mathcal K}\Upsilon p_0\TREV_{mn}=\tau p_0{\mathcal K}\Upsilon \TREV_{mn}=
p_0\tau{\mathcal K}\Upsilon \TREV_{mn}=p_0\hat T_{mn}$,
\hfill{(30.i)}
\vskip.7pc\noindent
$\hat T_{mn} p_j=\tau{\mathcal K}\Upsilon\TREV_{mn}p_j=-\tau {\mathcal K}\Upsilon p_j \TREV_{mn}=
\tau p_j{\mathcal K}\Upsilon \TREV_{mn}=p_j\tau{\mathcal K}\Upsilon \TREV_{mn}  p_j\hat T_{mn}$,
\hfill{(30.ii)}
\vskip.7pc\noindent
$\hat T_{mn} {\textsf j}_k=\tau{\mathcal K}\Upsilon\TREV_{mn}{\textsf j}_k=
-\tau{\mathcal K}\Upsilon {\textsf j}_k\TREV_{mn}= $\hfill{(30.iii)}\par\noindent
$\qquad\quad =-\tau{\mathcal K}\Upsilon\left(ip_l\frac{\partial}{\partial p_j}-ip_j\frac{\partial}{\partial p_l}\right)\TREV_{mn}
-\tau{\mathcal K}\Upsilon S_k\TREV_{mn}=$\par\noindent
$\qquad\quad=\tau\left(ip_l\frac{\partial}{\partial p_j}-ip_j\frac{\partial}{\partial p_l}\right){\mathcal K}\Upsilon\TREV_{mn}
-\tau\overline{S_k}\tau^{-1}\tau{\mathcal K}\Upsilon\TREV_{mn}=$
\vskip.5pc
$\qquad\quad=\left(ip_l\frac{\partial}{\partial p_j}-ip_j\frac{\partial}{\partial p_l}\right)\tau{\mathcal K}\Upsilon  \TREV_{mn}+ S_k\tau{\mathcal K}\Upsilon\TREV_{mn}
=  {\textsf j}_k\hat T_{mn}$,
\vskip.7pc\noindent
$\hat T_{mn}{\textsf k}_j=\tau{\mathcal K}\Upsilon \TREV_{mn}{\textsf k}_j=\tau{\mathcal K}\Upsilon {\textsf k}_j \TREV_{mn}=
\tau{\mathcal K}\Upsilon\left(ip_0\frac{\partial}{\partial p_j}-\frac{[{\bf S}\land{\bf p}]_j}{\mu+p_0}\right) \TREV_{mn}=$ \hfill{(30.iv)}
\vskip.5pc $\qquad\quad=\tau\left(ip_0\frac{\partial}{\partial p_j}+\frac{[\overline{\bf S}\land{\bf p}]_j}{\mu+p_0}\right){\mathcal K}\Upsilon\hat \TREV_{mn}=
\left(ip_0\frac{\partial}{\partial p_j}+\frac{[\tau\overline{\bf S}\tau^{-1}\land{\bf p}]_j}{\mu+p_0}\right)\tau{\mathcal K}\Upsilon\hat \TREV_{mn}=
{\textsf k}_j\hat T_{mn}$.
\vskip.8pc\noindent
The irreducibility of each component $U^{(m)}:{\mathcal P}_+^\uparrow\to{\mathcal U}(L_2(\RR^3,\CC^{2s+1},d\nu))$
implies that ${\hat T}=\left[\begin{array}{cc}d_{11}& d_{12} \cr d_{21} & d_{22}\end{array}\right]$, with $d_{mn}$ constant.
\vskip.8pc
Further constraints are imposed by the condition
$\TREV^2=c\Id$, with $c=\pm 1$. Now we have
$\TREV^2=\tau{\mathcal K}\Upsilon\hat T \tau{\mathcal K}\Upsilon\hat T=\tau\overline{\tau}\overline{\hat T}\hat T$;
therefore  $\TREV^2=\overline{\hat T}\hat T$ if $s\in\NN$ and $\TREV^2=-\overline{\hat T}\hat T$ if $s\in(\NN+\frac{1}{2})$.
It is clear that
there are always unitary constant matrices ${\hat T}=\left[\begin{array}{cc}d_{11}& d_{12} \cr d_{21} & d_{22}\end{array}\right]$ for which
$\TREV=(\tau{\mathcal K}\Upsilon\hat T)^2=\pm\Id$: it is sufficient that $\overline{\hat T}\hat T=\pm 1$; a trivial solution is ${\hat T}=\left[\begin{array}{cc}1& 0 \cr 0 & 1\end{array}\right]$,
that satisfies also (6) with the operator $\SREV$ given by (27);
but other less trivial solutions can easily singled out, such as $\hat T=\left[\begin{array}{cc}0& 1 \cr -1 & 0\end{array}\right]$.\par\noindent
So, extensions of $U^{(1)}\oplus U^{(2)}$ to generalized projective representations of the whole $\mathcal P$ are easily realized.
\vskip.5pc\noindent
{\bf Example 5.1.}
Let us study, for instance,
the case $s=0$, where $\tau=1$. The condition $\overline{\hat T}\hat T=\pm \Id$ entails ${\hat T}=c{\hat T}^t$,
where $c=\pm 1$ ant $\hat T^t$ being the transpose of $\hat T$.
\par
If $c=1$, then ${\hat T}={\hat T}^t$, that implies $\hat T=\left[\begin{array}{cc}d_{11}& d_{12} \cr d_{21} & d_{22}\end{array}\right]=
\left[\begin{array}{cc}d_{11}& d_{21} \cr d_{12} & d_{22}\end{array}\right]$, i.e. $d_{12}=d_{21}$.
Since ${\hat T}^2=\Id$ and $\hat T$ is unitary, $\hat T={\hat T}^{-1}={\hat T}^\ast$, i.e. $\hat T$ is hermitean and has two eigenvalues
$+1$ and $-1$. Therefore $\hat T={\bf n}\cdot{\vec\sigma}$.
Condition (6) implies $\omega=\pm 1$, and if $\omega=1$ then $\hat T=\left[\begin{array}{cc}0&1\cr 1&0\end{array}\right]$,
whereas if $\omega=-1$ then $\hat T=\left[\begin{array}{cc}1&0\cr 0&-1\end{array}\right]$.
\par
If $c=-1$, then ${\hat T}=c{\hat T}^t$ implies $\hat T=\left[\begin{array}{cc}0& d \cr -d & 0\end{array}\right]$, with $d\in\CC$. In this case ${\hat T}^2=-1$,
so that $\left[\begin{array}{cc}-d^2&0 \cr 0&-d^2 \end{array}\right]=\left[\begin{array}{cc}-1& 0 \cr 0 & -1\end{array}\right]$, i.e $d=\pm 1$ and
we can take $\hat T=\left[\begin{array}{cc}0& 1\cr -1 & 0\end{array}\right]$.
\par
The required commutation relation (6) between $\SREV$ and $\TREV$ is satisfied in case $c=-1$ with $\omega=-1$. Indeed,
$$
\SREV\TREV=\Upsilon{\mathcal K}\Upsilon\left[\begin{array}{cc}0& 1 \cr 1 & 0\end{array}\right]\left[\begin{array}{cc}0& 1 \cr -1 & 0\end{array}\right]
={\mathcal K}\left[\begin{array}{cc}-1& 0 \cr 0 & 1\end{array}\right],
$$
and
$$
\TREV\SREV={\mathcal K}\left[\begin{array}{cc}0& 1 \cr -1 & 0\end{array}\right]\left[\begin{array}{cc}0& 1 \cr 1 & 0\end{array}\right]
={\mathcal K}\left[\begin{array}{cc}1& 0 \cr 0 & -1\end{array}\right]=-\SREV\TREV.
$$
\subsubsection{Irreducibility of the extension $U:{\mathcal P}\to{\mathcal U}({\mathcal H})$.}
Now we show that, for each possible value of $s$,
there are irreducible generalized projective representation $U:{\mathcal P}\to{\mathcal V}({\mathcal H})$
that extend the reducible projective representations  $U:{\mathcal P}_+^\uparrow\to{\mathcal U}({\mathcal H})$
of the kind we are considering, that are irreducible.
\par
Let $A=\left[\begin{array}{cc}A_{11}& A_{12} \cr A_{21} & A_{22}\end{array}\right]$ be any self-adjoint operator of ${\mathcal H}=L_2(S_\mu^+,\CC^{2s+1},d\nu)\oplus L_2(S_\mu^+,\CC^{2s+1},d\nu)$, such that $[A,U_g]=\nop$ for all $g\in\mathcal P$, and therefore
$A$ commutes with all self-adjoint generators and with $\TREV$ and $\SREV$. From $[A,P_j]=\nop$ we imply that each $A_{mn}$ must be a function of $\bf p$:
$A_{mn}=a_{mn}({\bf p})$, and in particular $[A_{mn},p_j]=\nop$. Moreover, $[A, J_k]=\nop$ implies $[A_{mn},{\textsf j}_k]=\nop$.
Then, since $p_1,p_2,p_3,{\textsf j}_1,{\textsf j}_2,{\textsf j}_3$ are the generators of an irreducible projective representation of
$\mathcal E$ in the Hilbert space $L_2(S_\mu^+,\CC^{2s+1},d\nu)$, each $A_{mn}$ is a multiple of the identity: $A_{mn}=a_{mn}\Id$.
Now, the condition $[A,\SREV]=\nop$ implies $A=\left[\begin{array}{cc}a& b \cr b & a\end{array}\right]$.
In the generalized projective representation where $\hat T=\left[\begin{array}{cc}0& 1 \cr -1 & 0\end{array}\right]$
we have $\TREV=\tau\Upsilon{\mathcal K}\left[\begin{array}{cc}0& 1 \cr -1 & 0\end{array}\right]$;
the condition $[A,\TREV]=\nop$ implies $b=0$. Therefore, a self-adjoint operator $A$ that commutes with all $U_g$, $g\in{\mathcal P}$
must have the form $A=\left[\begin{array}{cc}a& 0 \cr 0 & a\end{array}\right]\equiv a\Id$, and therefore the generalized projective representation $U$ is irreducible.
\subsection{The case $\sigma(\underline P)=S_\mu^+\cup S_\mu^-$}
Now we determine irreducible representations $U$ of $\mathcal P$ with $\sigma(\underline P)=S_\mu^+\cup S_\mu^-$,
such that $U^+$, and hence $U^-$ by Prop. 4.1, is the direct sum of two irreducible projective representations $U^{(1)}$ and $U^{(2)}$ of ${\mathcal P}_+^\uparrow$.
Our search will be successful for $\TREV$ unitary and $\SREV$ anti-unitary.
\par
The aimed irreducibility forces the characterizing parameters  $\mu$ and $s$ of $U$ to have the same values for
the reduced components $U^{(1)}$ and $U^{(2)}$;
hence, $U^{(1)}$ and $U^{(2)}$ must be unitarily isomorphic, so that they can be identified with two identical projective representations according to section 4.2.1.
\par
We consider the case where $s=0$, because its simplicity helps clearness. Each of the Hilbert spaces ${\mathcal M}$ of $U^{(1)}$ and ${\mathcal N}$ of
$U^{(2)}$ can be identified with $L_2(\RR^3,d\nu)\oplus L_2(\RR^3,d\nu)$.
According to Prop. 4.1.i, both subspaces ${\mathcal M}={\mathcal M}^+\oplus {\mathcal M}^-$ and ${\mathcal N}={\mathcal N}^+\oplus {\mathcal N}^-$,
of ${\mathcal H}={\mathcal M}\oplus {\mathcal N}$,
where ${\mathcal M}^-=\TREV{\mathcal M}^+$ and ${\mathcal N}^-=\TREV{\mathcal N}^+$ reduce $U\mid_{{\mathcal P}_+}$. Hence, every vector $\psi$ of the Hilbert space
$\mathcal H$ of the entire generalized projective representation of $\mathcal P$ can be uniquely decomposed as
$\psi=\psi_{{\mathcal M}^+}+\psi_{{\mathcal M}^-}+\psi_{{\mathcal N}^+}+\psi_{{\mathcal N}^-}$, with
$\psi_{{\mathcal M}^+}\in{\mathcal M}^+$, $\psi_{{\mathcal M}^-}\in{\mathcal M}^-$,
$\psi_{{\mathcal N}^+}\in{\mathcal N}^+$, $\psi_{{\mathcal N}^-}\in{\mathcal N}^-$, so that $\psi$ can be represented as a column vector
$\psi=\left[\begin{array}{c}\psi_{{\mathcal M}^+}\cr\psi_{{\mathcal M}^-}\cr\psi_{{\mathcal N}^+}\cr\psi_{{\mathcal N}^-}\end{array}\right]$.
\par\noindent
In such a representation the self-adjoint generators of ${\mathcal P}_+^\uparrow$ satisfying (1) are
$$
P_0=\left[\begin{array}{cccc}p_0&0&0&0\cr 0&-p_0&0&0\cr 0&0&p_0&0\cr 0&0&0&-p_0\end{array}\right],\quad
P_j=\left[\begin{array}{cccc}p_j&0&0&0\cr 0&p_j&0&0\cr 0&0&p_j&0\cr 0&0&0&p_j\end{array}\right],
$$
$$
J_k=\left[\begin{array}{cccc}{\textsf j}_k&0&0&0\cr 0&{\textsf j}_k&0&0\cr 0&0&{\textsf j}_k&0\cr 0&0&0&{\textsf j}_k\end{array}\right],\quad
K_j=\left[\begin{array}{cccc}{\textsf k}_j&0&0&0\cr 0&-{\textsf k}_j&0&0\cr 0&0&{\textsf k}_j&0\cr 0&0&0&-{\textsf k}_j\end{array}\right]\,.
$$
According to Prop. 4.1, also the unitary operator $\TREV$ is reduced by $\mathcal M$ and $\mathcal N$, where its irreducible components,
by (21), are both $\left[\begin{array}{cc}0&1\cr 1&0\end{array}\right]$.
Then we have $\TREV=\left[\begin{array}{cccc}0&1&0&0\cr 1&0&0&0\cr 0&0&0&1\cr 0&0&1&0\end{array}\right]$.
\vskip.5pc\noindent
Now we seek for an anti-unitary space inversion operator $\SREV$.
\vskip.5pc
In the case $\SREV^2=\Id$, since $\SREV$ is anti-unitary, by imposing (3) we find
$\SREV={\mathcal K}\left[\begin{array}{cccc}0&s_1&0&s_2\cr s_1&0&s_2&0\cr 0&s_2&0&s_3\cr s_2&0&s_3&_0\end{array}\right]$,
with $s_1,s_2,s_3$ constant. So we have obtained a generalized projective representation of $\mathcal P$, because (1)-(6) hold.
\par
Such a representation, however, is reducible. Indeed
let $A=\left[\begin{array}{cccc}A_{11}&A_{12}&A_{13}&A_{14}\cr A_{21}&A_{22}&A_{23}&A_{24}\cr A_{31}&A_{32}&A_{33}&A_{34}\cr A_{41}&A_{42}&A_{43}&A_{44}
\end{array}\right]$ be any self-adjoint operator of $\mathcal H$;
the conditions $[A,P_0]=[A,P_j]=[A,J_k]=[A,K_j]=[A,\TREV]=[A,\SREV]=\nop$ are satisfied if and only if
$A=\left[\begin{array}{cccc}a&0&b&0\cr 0&a&0&b\cr \overline b&0&c&0\cr 0&\overline b&0&c
\end{array}\right]$ where $a,c\in\RR$ and $b\in\CC$, provided that $a+\overline b=b+c$.
Therefore, there are self-adjoint operators $A$ that commute with
all $U_g\in U({\mathcal P})$, different from a multiple of the identity. We have to conclude that if $\SREV^2=1$ then $U:{\mathcal P}\to{\mathcal V}({\mathcal H})$ is reducible.
\vskip.5pc
Let us now consider the case that $\SREV^2=-\Id$. We find that the conditions (3), (6) are satisfied if and only if
$\SREV={\mathcal K}\left[\begin{array}{cccc}0&0&0&1\cr 0&0&1&0\cr 0&-1&0&0\cr -1&0&0&0\end{array}\right]$.
If $A$ is any self-adjoint operator of $\mathcal H$, then this time the conditions
$[A,P_0]=[A,P_j]=[A,J_k]=[A,K_j]=[A,\TREV]=[A,\SREV]=\nop$ imply
$A=\left[\begin{array}{cccc}a&0&0&0\cr 0&a&0&0\cr 0&0&a&0\cr 0&0&0&a
\end{array}\right]\equiv a\Id$ with $a\in\RR$. Thus $U$ is irreducible.
\vskip.8pc
The results of sections 4 and 5 show that the whole class ${\mathcal I}_{\mathcal P}$ contains classes that are not considered in the literature about relativistic quantum theories of single particles;
for instance, in \cite{c13} only the representations of sections 4.1.1, 4.1.2 and
 $U^{(1)}$ and $U^{(2)}$ in section 4.3 are considered. Thus the present work identifies
two further (non-disjoint) robust classes of representations of $\mathcal P$ that should be considered for the formulation of relativistic quantum theories:
\vskip.5pc\noindent
${\mathcal I}_{\mathcal P}({\rm ant.}\SREV)$, i.e. the class that collects all representation of the kind $U^{(3)}$-$U^{(6)}$;
\vskip.5pc\noindent
${\mathcal I}_{\mathcal P}(U^\pm{\rm red.})$, i.e. the class of all representations in ${\mathcal I}_{\mathcal P}$ with $U^+$ or  $U^-$ reducible.
\section{Consistent relativistic quantum theories of elementary particle }
In the previous sections we have carried out a redetermination of the class of the irreducible generalized projective representations of $\mathcal P$, singling
out classes of irreducible representations besides those currently considered for the formulation of relativistic quantum theories of a particle.
Our work is meaningful, however, only if consistent theories based on these further representations can be developed.
This is the case, indeed; in this section some consistent theories of {\sl localizable particle} based on representations in the new classes,
derived in \cite{Gr32}, are presented.
\vskip.8pc
By {\sl localizable free particle}, shortly {\sl free particle} we mean
an isolated system whose quantum theory is endowed with a unique triple
$(Q_1,Q_2,Q_3)\equiv{\bf Q}$ of quantum observables, called {\sl position} operator, such that
\begin{itemize}
\item[({\it Q}.1)]
$[Q_j,Q_k]=\nop$, for all $j,k\in\{1,2,3\}$.
This condition requires that a measurement of position yields
all three values of the coordinates of the particle position.
\item[({\it Q}.2)]
The triple $(Q_1,Q_2,Q_3)\equiv{\bf Q}$ is characterized by
the specific properties of transformation of position with respect to the group $\mathcal P$,
expressed as relations for the transformed position observable $S_g[{\bf Q}]=U_g{\bf Q}U_g^{-1}$.
In particular,
\item[]\hskip-3pt(a)\quad
$S_\trev[{\bf Q}]={\bf Q}$ and $S_\srev[{\bf Q}]=-{\bf Q}$, equivalent to $\TREV{\bf Q}={\bf Q}\TREV$ and $\SREV{\bf Q}=-{\bf Q}\SREV$.
\item[]\hskip-3pt(b)\quad
If $g\in\mathcal E$ then $S_g[{\bf Q}]=U_g{\bf Q}U_g^{-1}={\textsf g}({\bf Q})$, where ${\bf x}\to{\textsf g}({\bf x})$ is the function that realizes $g$.
\end{itemize}
A free particle is said
{\sl elementary} if the generalized projective representation $U$ for which $S_g[A]=U_gAU_g^{-1}$ is irreducible.
Accordingly,
by selecting the irreducible generalized projective representations $U$
of $\mathcal P$, that admit such a triple $\bf Q$ satisfying ({\it Q}.1) and ({\it Q}.2) we identify the possible theories of elementary free particles. For the projective representations  with $\sigma(\underline P)=S_\mu^\pm$, $U^\pm$
irreducible and $s=0$, identified in section 4.1, it turns out that
conditions ({\it Q}.1) and ({\it Q}2.a,b) are sufficient \cite{Gr32} to univocally determine $\bf Q$
as $Q_j=F_j$, where $F_j=i\frac{\partial}{\partial p_j}-\frac{i}{2p_0^2}p_j$ are the Newton and Wigner operators \cite{N-W}. In this case, hence, we recover well known theories \cite{W2},\cite{JM},\cite{BM68}.
\subsection{Elementary particle theories with $s=0$ based on $U^{(3)}$ and $U^{(5)}$}
The explicit form of the tranformation properties with respect to $\mathcal P$ is available only for the subgroup generated by the Euclidean group $\mathcal E$
and $\{\srev,\trev\}$; they are expressed by ({\it Q}.2,a,b).
For the irreducible generalized projective representations with $\sigma(\underline P)=S_\mu^+\cup S_\mu^-$, $U^\pm$
irreducible and $s=0$, identified in section 4.2, the known transformation properties
({\it Q}.1) and ({\it Q}2.a,b) are sufficient \cite{Gr32} to completely and univocally determine $\bf Q$ only for $U^{(3)}$ and $U^{(5)}$; the position operator must be
${\bf Q}=\hat{\bf F}=\left[\begin{array}{cc}F_j&0\cr 0&F_j\end{array}\right]$.
Hence, we have two complete theories based on the new representations $U^{(3)}$ and $U^{(5)}$.
Though in $U^{(3)}$ the space inversion operator is anti-unitary, and in $U^{(5)}$ also the time reversal operator is anti-unitary, the theories are perfectly consistent, in the sense that ({\it Q}.1) and ({\it Q}.2) are satisfied.
Thus, these new representations are indispensable to determine complete theories with the nowadays available conditions.
\vskip.5pc
The early theory for such a kind of particle is Klein-Gordon theory \cite{Klein}-\cite{Gordon}, that suffered serious problems.
A first problem is that the wave equation of Klein-Gordon theory is second order in time, while according to the general laws of quantum theory it should be first order.
\par
Furthermore, Klein-Gordon theory interprets
$\hat\rho(t,{\bf x})=\frac{i}{2m}\left(\overline{\psi_t}\frac{\partial}{\partial t}\psi_t-{\psi_t}\frac{\partial}{\partial t}\overline{\psi_t}\right)$
as the {\sl probability of position} density and
$\hat{\bf j}(t,{\bf x})=\frac{i}{2m}\left(\psi_t\nabla\overline{\psi_t}-\overline{\psi_t}\nabla\psi_t\right)$ as
its {\sl current} density.
This interpretation is at the basis of the Dirac concern that position probability density can be negative, due to the presence of time derivatives of $\psi_t$ in $\hat\rho$.
A way to overcome the difficulty without making resort to quantum field theory \cite{PW} was proposed by Feshbach and Villars \cite{FV}.
They derive an equivalent form of Klein-Gordon equation as a first order equation
$i\frac{\partial}{\partial t}\Psi_t=H\Psi_t$
for the state vector $\Psi_t=\left[\begin{array}{c}\phi_t\cr \chi_t\end{array}\right]$, where
$\phi_t=\frac{1}{\sqrt{2}}(\psi_t+\frac{1}{m}\frac{\partial}{\partial t}\psi_t)$,
$\chi_t=\frac{1}{\sqrt{2}}(\psi_t-\frac{1}{m}\frac{\partial}{\partial t}\psi_t)$, and
$H=(\sigma_3+\sigma_2)\frac{1}{2m}(\nabla+m\sigma_3)$, $\psi_t$ being the Klein-Gordon wave function;
in this representation $\hat\rho=\vert\phi_t\vert^2-\vert\chi_t\vert^2$, without time derivatives.
The minus sign in $\hat\rho$ forbids to interpret it as probability density of position;
Feshbach and Villars proposed to reinterpret it as {\sl density probability of charge}, so that negative values could be accepted.
Nevertheless, according to Barut and Malin \cite{BM68}, covariance with respect to boosts should imply that
$\hat\rho$ must be the time component of a four-vector.
Barut and Malin proved that is not the case.
\vskip.5pc
In order to check our theories with respect to these problems, we reformulate the theories based on $U^{(3)}$ and $U^{(5)}$ in
equivalent forms, obtained by means of unitary transformations
operated by the unitary operator $Z=Z_1Z_2$, where $Z_2=\frac{1}{\sqrt{p_0}}\Id$ and $Z_1$ is the inverse of the {\sl Fourier-Plancherel} operator, that transforms $\psi({\bf p})$ into $(Z\psi)({\bf x})\equiv(\hat\psi)({\bf x})$.
In the so reformulated theories the Hilbert space for both turns out to be
${\mathcal H}=Z\left(L_2(\RR^3,d\nu)\oplus L_2(\RR^3,d\nu)\right)\equiv L_2(\RR^3)\oplus L_2(\RR^3)$; the new self-adjoint generators are
${\hat P}_j=ZP_jZ^{-1}=\left[\begin{array}{cc}-i\frac{\partial}{\partial x_j}&0\cr 0&-i\frac{\partial}{\partial x_j}\end{array}\right]$,
${\hat P}_0=\sqrt{\mu^2-\nabla^2}\left[\begin{array}{cc}1&0\cr 0&-1\end{array}\right]$,
${\hat J}_k=-i\left(x_l\frac{\partial}{\partial x_j}-x_j\frac{\partial}{\partial x_l}\right)\left[\begin{array}{cc}1&0\cr 0& 1\end{array}\right]$;
${\hat K}_j=\frac{1}{2}\left(x_j\sqrt{\mu^2-\nabla^2}+\sqrt{\mu^2-\nabla^2}x_j\right)\left[\begin{array}{cc}1&0\cr 0&-1\end{array}\right]$.
\par\noindent
The wave equation trivially is $i\frac{\partial}{\partial t}\psi_t=P_0\psi_t$, that is first order.\par
The position operator turns out to be
${\hat Q}_j=Z\hat{\bf F}Z^{-1}\equiv\left[\begin{array}{cc}x_j&0\cr 0&x_j\end{array}\right]$, so that also the other problems disappear.
Indeed, the position is represented by the multiplication operator;
therefore,
the probability density of position must necessarily be given by the non negative function
$\rho(t,{\bf x})=\vert\psi_t^+({\bf x})\vert^2+\vert\psi_t^-({\bf x})\vert^2$.
On the other hand,  being $K_j$ and $\bf Q$ explicitly known,
the covariance properties with respect to boosts, according to ({\it Q}.2), are explicitly expressed in full coherence by
$S_g[{\bf Q}]=e^{iK_j\varphi(u)}{\bf Q}e^{-iK_j\varphi(u)}$.
\subsection{New species of particle theories}
In the literature all irreducible representations taken as bases of elementary particle theories
are characterized by the irreducibility of $U^\pm$.
Now, in section 5.1 for each $\mu>0$ and every $s\in\frac{1}{2}\NN$ an irreducible representation of $\mathcal P$ is identified characterized by
$\sigma(\underline P)=S_\mu^+$ such that
$U^+$ is {\sl reducible}. It can be shown \cite{Gr32} that
conditions ({\it Q}.1), ({\it Q}.2.a,b) univocally determine the position operator $\hat{\bf Q}$, and therefore gives rise to a consistent theory \cite{Gr32}.
For these representations,
where ${\mathcal H}=L_2(\RR^3,d\nu)\oplus L_2(\RR^3,d\nu)$, such position operator is ${\hat Q}_j=\left[\begin{array}{cc}F_j&0\cr 0&F_j\end{array}\right]$.
Therefore, complete consistent theories of an elementary free particle turn out to be identified, which corresponds to none of the early theories.
\par
Thus, the extension of the class of the irreducible representations of $\mathcal P$ is meaningful, because it allows to identify consistent theories and also
new species of consistent theories.

\end{document}